\documentclass[sigplan,10pt]{acmart}

\renewcommand\footnotetextcopyrightpermission[1]{}
\setcopyright{none}
\usepackage{amsmath}

\usepackage{filecontents}
\usepackage[normalem]{ulem}
\usepackage{graphicx}
\usepackage{enumitem}
\usepackage[linesnumbered,ruled,vlined]{algorithm2e}
\usepackage{algpseudocode}
\usepackage{tikz}
\usepackage{pifont}
\usepackage{xcolor}
\usepackage{cleveref}
\usepackage{listings}

\crefname{section}{§}{§§}
\Crefname{section}{§}{§§}

\usepackage{amsthm}

\hypersetup{
    colorlinks=true,
    linkcolor=blue,
    citecolor=blue,
    urlcolor=blue
}

\lstdefinestyle{cpp}{
  language=C++,
  basicstyle=\ttfamily\footnotesize,
  keywordstyle=\color{blue}\bfseries,
  commentstyle=\color{gray}\itshape,
  stringstyle=\color{red},
  numbers=none,
  breaklines=true,
  keepspaces=true,
  columns=fullflexible,
  aboveskip=3pt,
  belowskip=3pt,
  showstringspaces=false,
  tabsize=2
}
\newcommand{\fref}[1]{Fig.~\textcolor{blue}{\ref{#1}}}
\newcommand{\tref}[1]{Table~\textcolor{blue}{\ref{#1}}}

\newcommand{\sref}[1]{Step~\textcolor{blue}{\ref{#1}}}

\graphicspath{{Figures/}{../}}

\newcommand{\DEL}[1]{\iffalse #1 \fi}

\begin{document}

\title{C2CServe: Leveraging NVLink-C2C for Elastic Serverless LLM Serving on MIG}

\author{Shutian Luo}
\affiliation{
  \institution{University of Virginia}
  \country{USA}
}

\author{Ali Zafar Sadiq}
\affiliation{
  \institution{University of Virginia}
  \country{USA}
}

\author{Rui Yang}
\affiliation{
  \institution{University of Virginia}
  \country{USA}
}

\author{Mingye Zhang}
\affiliation{
  \institution{Northwestern University}
  \country{USA}
}

\author{Haiying Shen}
\affiliation{
  \institution{University of Virginia}
  \country{USA}
}

\author{Wei Wang}
\affiliation{
  \institution{HKUST}
  \country{Hong Kong SAR, China}
}

\author{Yue Cheng}
\affiliation{
  \institution{University of Virginia}
  \country{USA}
}

\begin{abstract}
Modern LLM serving is increasingly serverless in shape: large model catalogs,
long-tail invocations, and multi-tenant demand. Existing GPU serving systems face
a tradeoff:  dedicated-GPU allocation wastes scarce HBM under sparse traffic,
while GPU time sharing places model initialization and weight loading on the
cold-start path. Spatial GPU sharing such as multi-instance GPU (MIG) provides isolation and
accounting, but each slice has too little HBM for modern LLM weights.

    We observe that high-bandwidth CPU--GPU interconnects, such as NVLink-C2C (C2C) in NVIDIA GH200 and GB200 Superchips, change the memory constraint: model weights can reside in CPU memory and be streamed on demand to MIG instances, shifting model residency from scarce HBM to abundant host memory. 
    Leveraging this capability, we present C2CServe, a request-granularity serverless LLM serving system that allows MIG instances to switch models across requests without reloading weights into HBM.
    C2CServe introduces \emph{HybridGEMM}, a heterogeneous-memory-aware GEMM kernel that adapts data access patterns to balance HBM and C2C bandwidth across MIG partitions using a single tuning knob. To mitigate shared-C2C contention, C2CServe further uses a hierarchical scheduler that coordinates model placement, input chunking, and kernel selection with online feedback control. 
    On GH200, C2CServe reduces cold-start latency by up to 7.1\(\times\) for dense models and 4.6\(\times\) for MoE models compared with state-of-the-art serverless LLM serving systems, while maintaining over 95\% TTFT and TPOT attainment under C2C contention.
    
\end{abstract}

\maketitle

\section{Introduction}
\label{sec:intro}
Large language model serving~\cite{kaplan2020scaling, achiam2023gpt,touvron2023llama,vaswani2017attention} is shifting from a small number of static deployments to large catalogs of dynamically invoked model variants. A single cloud tenant may register domain-specialized fine-tunes~\cite{sheng2023s}, periodically refreshed checkpoints~\cite{wan2025bytecheckpoint}, and request-routed expert mixtures behind a unified inference endpoint~\cite{moe-infinity}. Public model hubs already catalog over a million models~\cite{huggingfaceDataset}, and production traces~\cite{GenAI,lin2025understanding} from large-scale inference platforms show a pronounced long tail (detailed in~\cref{sec:motiv_workload}): a small fraction of models receives most requests, while the remaining models must still remain responsive to unpredictable invocations~\cite{lin2025understanding} (details in~\cref{sec:motiv_workload}). This long-tail workload closely matches the serverless setting---long periods of inactivity punctuated by bursty, multi-tenant invocations~\cite{fu2024serverlessllm,mao2025skyserve,zeng2025medusa,xiang2025aegaeon}. 

Serving long-tail workloads requires high elasticity: the system allocates resources at fine granularity for resource efficiency
while keeping cold-start overhead low. Yet both requirements conflict with the characteristics of modern LLMs. Their large model footprints force coarse-grained placement, often requiring one or more GPUs per model due to limited HBM capacity. Meanwhile, LLM cold starts go beyond image loading and container startup: inference engines must also initialize runtime state, and construct CUDA graphs~\cite{zeng2025medusa}. As model sizes continue to grow, high elasticity becomes increasingly difficult.

\begin{figure}[tbp]	\centerline{\includegraphics[width=0.98\linewidth]{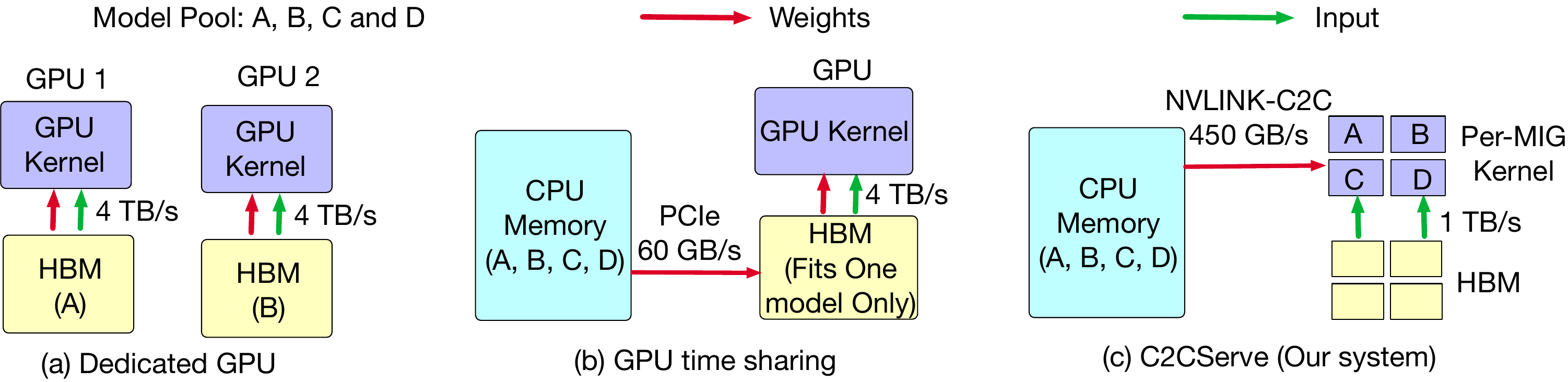}}
	\vspace{-0.5em}
 \caption{Multi-model serving approaches.}
           \vspace{-0.15in}
	\label{fig:Multi-model-serving}
\end{figure}

Existing GPU-based serving systems struggle to provide both fine-grained allocation and low cold-start overhead. \emph{Dedicated-GPU allocation} assigns one or more GPUs to each model~\cite{hu2025deepserve,griggs2024m}, as shown in~\fref{fig:Multi-model-serving}(a). This keeps models warm and avoids cold starts, but wastes accelerator memory and compute when traffic is sparse. 
In contrast, \emph{GPU time sharing} multiplexes multiple models on the same GPU through time sharing~\cite{xiang2025aegaeon,yu2026taming,he2025resource,choi2022serving,li2022tetris,duan2024muxserve}, as shown in~\fref{fig:Multi-model-serving}(b). This improves utilization, but only the active model can be served at a time. When the active model changes, the system must initialize inference-engine state and load gigabytes of weights over PCIe, adding substantial overhead on the cold-start path~\cite{zeng2025medusa}.
Unlike conventional serverless functions, LLM cold starts are dominated by heavyweight GPU-side setup and weight materialization rather than lightweight container startup.

This tension suggests the need for an allocation unit that is finer-grained than a full GPU, yet stable enough to keep serving state warm across requests. NVIDIA Multi-Instance GPU (MIG) appears to offer such a middle ground for serverless LLM serving. 
MIG can partition a single accelerator into up to seven isolated instances~\cite{MIGs}, each with dedicated compute resources and a private fraction of HBM. This makes MIG attractive for multi-tenant inference~\cite{vijaykumar2016zorua,strati2024orion,han2022microsecond,he2025resource}: each instance can serve a separate model, reduce model-switch frequency to mitigate cold-start requirement, and provide a natural per-instance accounting unit, as shown in~\fref{fig:Multi-model-serving}(c). However, its scalability is limited by partitioned HBM capacity: modern LLMs already strain full-GPU memory, and MIG further divides HBM across instances. As a result, each slice is often too small to keep the model weights it is expected to serve resident in HBM. MIG therefore provides the appropriate execution and accounting abstraction, but its memory granularity is ill-suited for modern LLM footprints.

This limitation motivates using CPU memory as the weight store for MIG instances. Integrated CPU--GPU accelerators alleviate the MIG memory bottleneck by coupling GPU compute with high-bandwidth host memory. NVIDIA Superchips such as GH200~\cite{GH200} and GB200~\cite{GB200} provide up to \(\sim\)450\,GB/s per-direction NVLink-C2C (C2C) bandwidth, about \(7\times\) higher than PCIe 5.0 x16 (\(\sim\)64\,GB/s per direction). Existing systems~\cite{xu2024pie,yu2026superinfer,li2025oneiros} treat CPU memory as a faster HBM backing tier, similar in spirit to PCIe-based offloading. In contrast, we observe that C2C enables CPU memory to act as an active extension of HBM: model weights can remain in host memory~\cite{PinnedMemory, cudahostalloc} and be consumed {directly} by GPU kernels via zore-copy access~\cite{ZeroCopy}, without being staged into HBM.

\emph{Together, MIG and C2C make LLM serverless practical:} MIG provides fine-grained compute, while C2C extends each MIG instance beyond its private HBM partition to a larger CPU-memory weight store. We observe that their combination changes the cost structure of serverless LLM serving as shown in~\fref{fig:Multi-model-serving}(c): cold starts no longer require copying model weights into HBM, and inactive models no longer occupy scarce GPU memory. Instead, long-tail models can reside in CPU memory and be streamed on demand by any MIG instance. This shifts the admission bottleneck from HBM capacity to host-memory capacity, enabling substantially higher model density while reserving HBM for active per-request state such as activations and KV cache.

Realizing this design requires rethinking two assumptions in today’s GPU software stack. 
First, existing general matrix multiplication (GEMM) kernels such as cuBLAS~\cite{cublas} and CUTLASS~\cite{CUTLASS} assume HBM-resident operands. When directly used with CPU-resident weights, they can repeatedly fetch the same weight blocks over NVLink-C2C, amplifying C2C traffic. This mismatch is more pronounced under MIG, where HBM bandwidth is partitioned across instances but C2C bandwidth remains shared.
Second, C2C sharing weakens MIG isolation. Co-resident MIG instances may stream CPU-resident weights concurrently, so each tenant’s effective C2C bandwidth depends on aggregate demand rather than its own partition.
This creates a scheduling challenge: coordinating shared C2C bandwidth with partitioned MIG-local compute and HBM resources.

We present C2CServe, a request-granularity serverless LLM serving system that decouples model residency from GPU HBM. By keeping model weights in CPU memory and streaming them on demand over C2C without staging them in HBM, C2CServe allows each MIG instance to switch models at request granularity.

Specifically, C2CServe introduces \emph{HybridGEMM} with key insight to trade C2C traffic for HBM traffic. HybridGEMM splits execution between an output-stationary path that preserves GEMM efficiency and a weight-stationary path that reuses CPU-resident weights to reduce repeated C2C fetches. A single knob adapts this tradeoff to each MIG slice's HBM--C2C balance and runtime contention. On top of HybridGEMM, {C2CServe} designs a C2C-aware scheduler for serverless LLM serving. For each request, the scheduler considers the model size, selects the input chunk size, and chooses a precompiled HybridGEMM variant whose HBM--C2C ratio matches the slice's live bandwidth profile. At runtime, C2CServe monitors serving performance and further tunes the HybridGEMM knob through feedback control, enabling efficient support for long-tail models with diverse memory footprints and bursty traffic.

We evaluate C2CServe on GH200 Superchips using multi-tenant production workloads across model sizes. 
By eliminating expensive weight loading, C2CServe reduces cold-start latency by up to 7.1\(\times\) on dense models and 4.6\(\times\) on MoE models. Even under multi-tenant C2C contention, C2CServe maintains over 95\% attainment for both Time-To-First-Token (TTFT) and Time-Per-Output-Token (TPOT) targets.
This paper makes three contributions.
\begin{itemize}
\item We identify the opportunity for serverless LLM inference on Superchips: MIG provides fine-grained compute isolation, while C2C enables CPU-resident model weights. To our knowledge, C2CServe is the first system to make MIG practical for serverless LLM serving by decoupling model residency from scarce HBM capacity.

\item We design HybridGEMM, a heterogeneous-memory-aware GEMM kernel that adapts to each MIG slice's HBM/C2C bandwidth profile through a single tuning knob, rather than relying on one fixed tiling strategy.

\item We design a hierarchical scheduling policy across model placement, input chunking, and kernel selection to mitigate C2C contention in multi-tenant serverless serving.
\end{itemize}
\section{Background}
\label{sec:background}

\subsection{LLM Workload Characterization}
\label{sec:motiv_workload}
\begin{figure}[tbp]	\centerline{\includegraphics[width=0.98\linewidth]{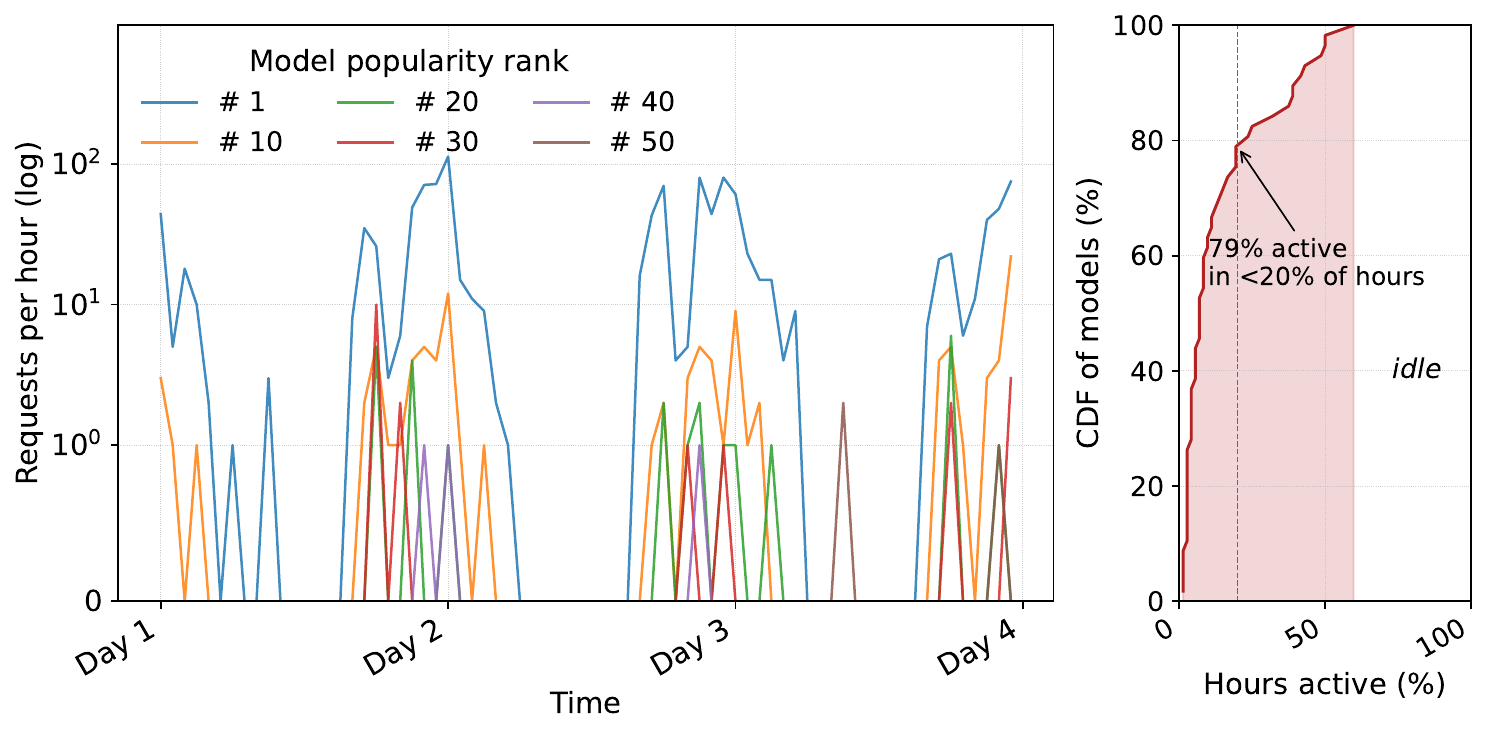}}
	\vspace{-0.5em}
 \caption{
LLM workload fluctuation in an Alibaba production cluster.
\textbf{Left:} Hourly request rates of representative models.
\textbf{Right:} Per-model active-time distribution across 59 active models.
}
           \vspace{-0.15in}
    \label{fig:workload_trace}
\end{figure}

We characterize LLM serving behavior using a three-week Alibaba production trace~\cite{lin2025understanding,GenAI} containing 89 models as shown in ~\fref{fig:workload_trace}. Three observations stand out. First, per-model traffic is highly bursty: requests arrive in short bursts separated by long idle periods, and the median active model is idle for 96\% of observed hours as shown in~\fref{fig:workload_trace}(a). Second, bursts are not perfectly aligned across models; one model's peak often coincides with another model's idle period, making aggregate demand smoother than individual demand. Third, activity is strongly long-tailed: 83\% of models are active in fewer than 20\% of observed hours as depicted in~\fref{fig:workload_trace}(b), yet these models must still remain responsive to unpredictable requests. These properties match the canonical serverless workload pattern and expose a structural mismatch with existing GPU serving systems.

\subsection{Elastic LLM Serving under Long-Tail Demand}
Serverless LLM serving targets long-tail workload scenarios~\cite{lin2026flexpipe} where models may remain inactive for long periods but must respond quickly when requests arrive. To achieve high elasticity, existing systems commonly rely on parameter offloading and model switching. Parameter offloading moves model weights out of scarce GPU HBM, typically to CPU memory or storage, and loads them back when the model is invoked. Model switching allows multiple models to share the same GPU over time, improving utilization under sparse demand.

However, these mechanisms also introduce new overheads. When a request arrives for an inactive model, the system must reload model weights, initialize inference-engine state, and prepare GPU execution resources before serving the request. As a result, serverless LLM systems must balance two competing goals: supporting many rarely used models with limited GPU memory, while keeping cold-start latency low enough for interactive serving.

\subsection{MIG Partitioning}
\label{sec:bg_mig}
NVIDIA MIG technology~\cite{MIGs}, introduced with A100 and extended to H100 and GH200, partitions a GPU into up to seven hardware-isolated instances. Each instance receives dedicated GPU resources, including HBM capacity, HBM bandwidth and SMs. As summarized in~\tref{tab:gh200_mig_config}, GH200 supports configurations ranging from one full-GPU instance with 96\,GB HBM and 132 SMs to seven small instances, each with 12\,GB HBM and 16 SMs.

This partitioning exposes a fundamental tradeoff: higher instance concurrency comes at the cost of lower per-instance compute capacity, HBM capacity, and HBM bandwidth. While this makes MIG attractive for isolation, accounting, and multi-tenant serving, it also tightly bounds each instance by its private HBM resources. This limitation is particularly severe for LLM inference: a 70B-parameter model in BF16 requires roughly 140\,GB for weights alone, exceeding even the 96\,GB HBM capacity of a full GH200 GPU and far exceeding the capacity of smaller MIG instances. As a result, serving large models on MIG requires quantization, tensor-parallel sharding, or offloading, leaving MIG underused for production LLM serving despite its natural fit for fine-grained GPU partitioning.

\subsection{Superchip Memory Hierarchy}
\label{sec:bg_superchip}
NVIDIA GH200~\cite{GH200} and GB200~\cite{GB200} Superchips integrate a Grace CPU with a Hopper GPU and a Blackwell GPU, respectively, over C2C. This exposes a heterogeneous memory hierarchy: high-bandwidth GPU HBM for latency-critical state, and a larger CPU memory pool accessible to GPU kernels through the high-bandwidth C2C interconnect.

This hierarchy differs from a discrete-GPU system in both capacity and bandwidth. First, CPU memory provides a much larger and cheaper storage tier that can hold many model snapshots that would not fit in HBM, either individually or collectively. Second, C2C provides enough bandwidth for GPU kernels to read CPU-resident operands directly, without first staging them through HBM. As a result, CPU memory can act as an active operand store rather than merely a slow backing store. This changes model serving from an HBM-residency problem into a heterogeneous-memory scheduling problem: model weights can remain in CPU memory,
while scarce HBM is reserved for latency-critical per-request state.

\begin{table}[t]
\centering
\scriptsize
\setlength{\tabcolsep}{3pt}
\caption{GH200 MIG configs: [instances, HBM per instance], SM count, and partitioned HBM bandwidth.}
\label{tab:gh200_mig_config}
\begin{tabular}{c|ccccc}
\toprule
Config. & [1, 96GB] & [2, 48GB] & [3, 24GB] & [4, 24GB] & [7, 12GB] \\
\midrule
SMs per instance & 132 & 56 & 28 & 16 & 16 \\
HBM BW per instance & 4.0\,TB/s & 2.0\,TB/s & 1.0\,TB/s & 1.0\,TB/s & 0.5\,TB/s \\
\bottomrule
\end{tabular}
\vspace{-1.0em}
\end{table}

\section{Motivation for C2CServe}
\subsection{Opportunity of Combining MIG and C2C}
Serverless LLM serving requires high elasticity: low cold-start latency and fine-grained resource allocation. However, both are difficult when model weights must remain resident in scarce HBM.
High-bandwidth CPU--GPU interconnects such as C2C change this tradeoff. MIG provides lightweight, isolated GPU slices for multi-tenant execution and accounting, while C2C allows GPU kernels to stream CPU-resident model weights at high bandwidth. Thus, model parameters can remain in CPU memory, and each MIG instance can reserve its limited HBM for active execution state such as activations and KV cache. 

\subsection{GEMM on MIG-Partitioned Superchips}
Efficiently combining MIG with Superchip memory requires GPU kernels that are aware of operand placement and bandwidth asymmetry across the memory hierarchy. We use GEMM as a representative kernel to illustrate our MIG-specific kernel design.

\subsubsection{Symmetric GEMM and Asymmetric GEMM}
\begin{figure}[tbp]	\centerline{\includegraphics[width=0.95\linewidth]{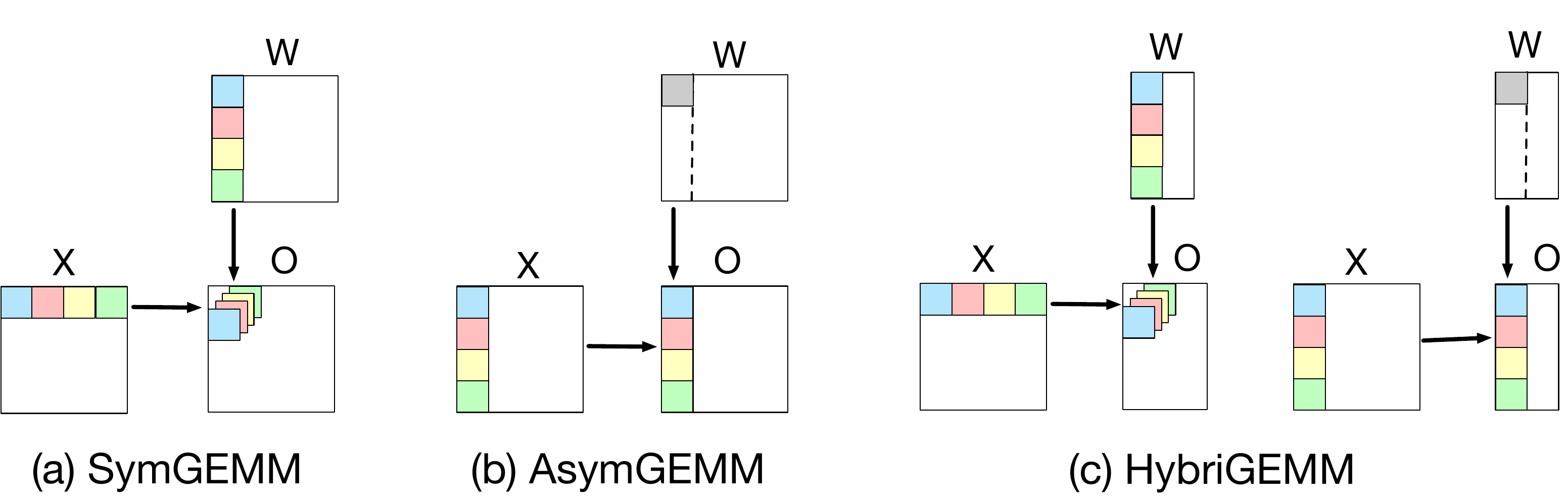}}
	\vspace{-0.5em}
    \caption{Comparison of data access patterns across different GEMM tiling strategies.}
    \vspace{-0.15in}
	\label{fig:HybridGEMM}
\end{figure}

Given a GEMM operation \(O = X \times W\), conventional kernels use a \emph{symmetric GEMM} (\emph{SymGEMM}) dataflow that treats \(X\) and \(W\) with roughly balanced access cost, as shown in~\fref{fig:HybridGEMM}(a). In contrast, \emph{asymmetric GEMM} (\emph{AsymGEMM}) keeps \(W\) stationary and streams \(X\) and \(O\), reducing repeated accesses to CPU-resident weights at the cost of additional HBM traffic for output accumulation, as shown in~\fref{fig:HybridGEMM}(b). Since the HBM-to-C2C bandwidth ratio varies across MIG partitions, neither dataflow is always optimal. We therefore introduce \emph{HybridGEMM}, which partitions SMs between \emph{SymGEMM} and \emph{AsymGEMM} to adapt the aggregate data movement to the available HBM and C2C bandwidth as shown in~\fref{fig:HybridGEMM}(c) and detailed in~\cref{sec:HybridGEMM-Kernel}.

\begin{figure}[tbp]
  \centering
\includegraphics[width=0.43\textwidth]{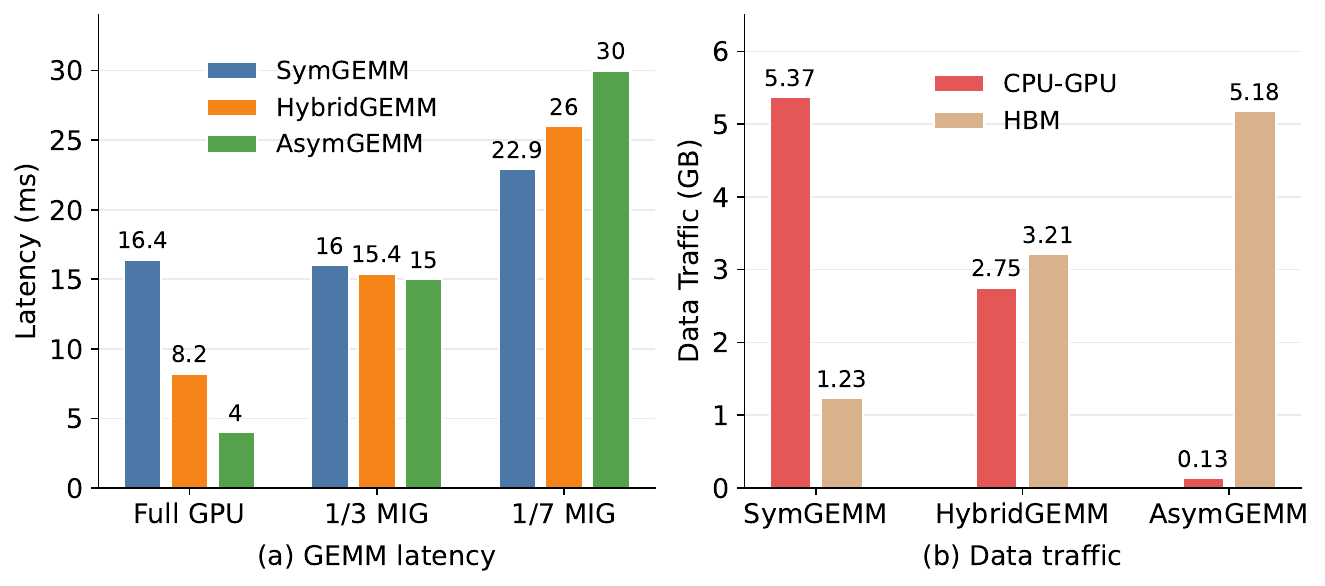}
\vspace{-.8em}
\caption{Comparison of symGEMM, asymGEMM, and HybridGEMM on a representative LLM inference GEMM shape (\(A: 10240 \times 4096\), \(B: 16384 \times 4096\)).}
\label{fig:different_GEMM}
\end{figure}

We evaluate a representative LLM-inference GEMM where activations reside in HBM and parameters reside in CPU memory. As shown in~\fref{fig:different_GEMM}(a), \emph{AsymGEMM} is highly effective on a full GPU, reducing latency from 16.4\,ms to 4.0\,ms by cutting CPU--GPU traffic. Under MIG, however, this advantage diminishes: \emph{SymGEMM} and \emph{AsymGEMM} perform similarly with 3 MIG instances, while \emph{AsymGEMM} becomes slower with 7 instances. This is because MIG partitions HBM bandwidth across instances while C2C bandwidth remains shared, narrowing the effective HBM-over-C2C bandwidth advantage.

Figure\textcolor{blue}{~\ref{fig:different_GEMM}}(b) explains this shift: \emph{AsymGEMM} reduces CPU--GPU transfers from 5.37\,GB to 0.13\,GB, but increases HBM traffic from 1.23\,GB to 5.18\,GB. Thus, \emph{SymGEMM} stresses C2C by repeatedly streaming CPU-resident weights, whereas \emph{AsymGEMM} shifts traffic to HBM by reusing weights and accumulating output updates there. \emph{HybridGEMM} interpolates between these extremes by adjusting the fraction of SMs assigned to each dataflow, making GEMM on Superchip systems bandwidth-adaptive rather than fixed.

Overall, these results suggest that GEMM on superchip systems should be bandwidth-adaptive rather than fixed. HybridGEMM provides such adaptability by tuning the balance between HBM traffic and CPU--GPU traffic to match the effective resource regime under each MIG configuration.

\begin{figure}[tbp]
  \centering
\includegraphics[width=0.43\textwidth]{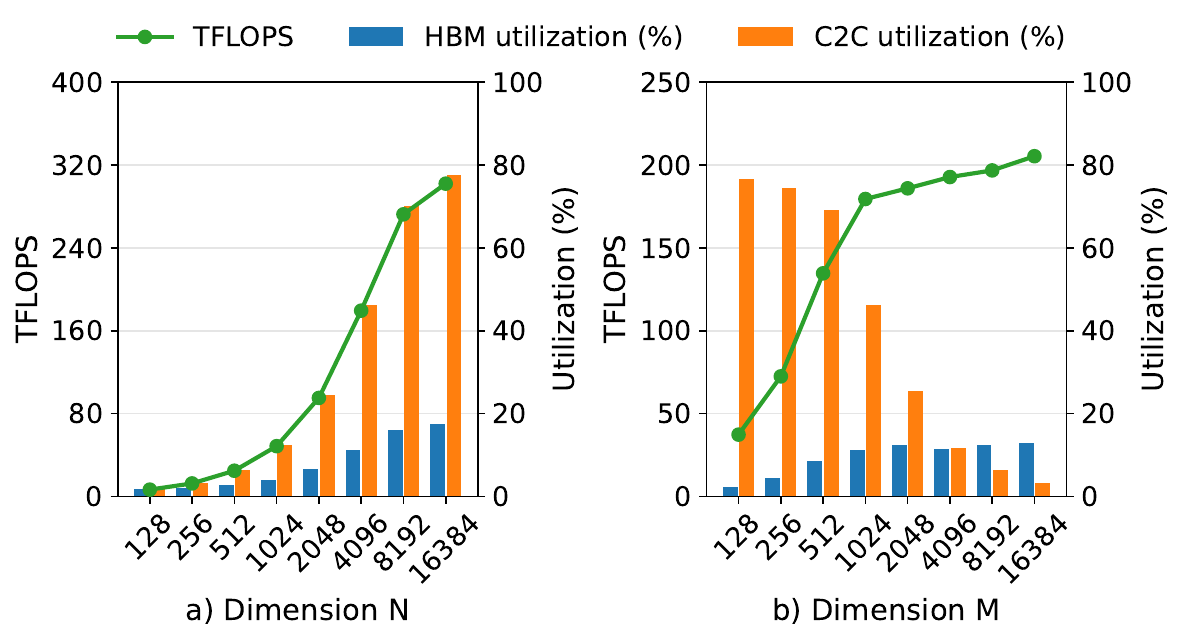}
\vspace{-.8em}
  \caption{Shape-dependent performance and bandwidth utilization on asymGEMM.}
\label{fig:shapeGEMM}
\end{figure}

\subsubsection{Shape-dependent Bottlenecks}
Directly accessing CPU-resident parameters is efficient only when the GEMM shape provides enough reuse to amortize CPU--GPU traffic. We study GEMMs \(O_{M \times N}=X_{M \times K} \times W_{K \times N}\), where \(X\) resides in HBM and \(W\) resides in CPU memory. Because the two operands come from different memory domains, matrix shape determines both HBM traffic and C2C pressure.

Figure~\ref{fig:shapeGEMM}
shows two distinct effects. Increasing \(N\) improves TFLOPS but sharply increases C2C utilization, pushing execution toward a C2C-bound regime. In contrast, increasing \(M\) improves TFLOPS while reducing C2C utilization, because more activation rows reuse the same parameter tiles and better amortize CPU-memory fetches. Thus, \(N\) primarily increases interconnect pressure, whereas \(M\) improves compute efficiency through higher parameter reuse. These results reveal a shape-dependent bottleneck shift: small shapes underutilize the GPU, large \(N\) makes execution C2C-bound, and large \(M\) makes direct CPU-parameter access more efficient by increasing reuse.

{\textbf{Takeaway.}} Superchip GEMM performance is bottleneck-dependent: MIG partitioning changes the effective HBM--C2C bandwidth balance, while matrix shape determines how well C2C traffic is amortized. High performance therefore requires an adaptive GEMM dataflow that jointly manages HBM and C2C traffic.

\subsection{Cross-Instance C2C Contention}
Although CPU offloading makes MIG-based multi-model serving practical, it also introduces a new source of interference: the shared CPU--GPU interconnect. In C2CServe, different MIG instances are isolated in on-chip compute resources and HBM allocation, but they still share the C2C path used to fetch CPU-resident model parameters. As a result, MIG isolation is only partial: computation is spatially partitioned on the GPU, whereas parameter access remains a shared off-chip resource. When multiple instances issue parameter fetches concurrently, contention on C2C can become a first-order performance bottleneck.

To understand this interference, we study two factors that directly shape C2C demand: \emph{parameter footprint} and \emph{execution granularity}. The former captures the volume of model data fetched from CPU memory, while the latter reflects how aggressively each instance utilizes the shared interconnect during execution.

\begin{figure}[tbp]
  \centering
\includegraphics[width=0.43\textwidth]{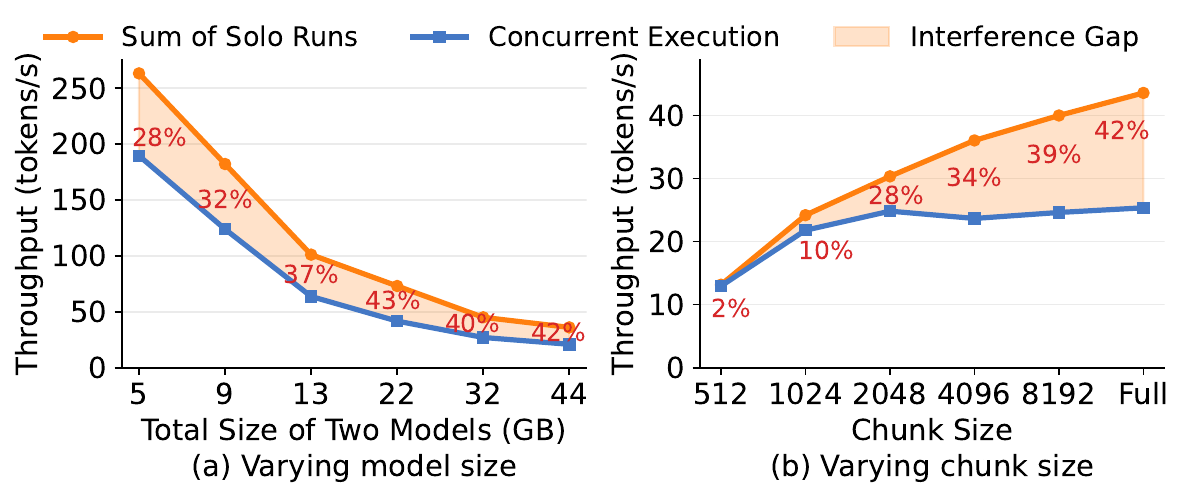}
\vspace{-.8em}
\caption{Interference on shared C2C bandwidth.}
\label{fig:shapeGEMM}
\end{figure}

\subsubsection{Impact of Parameter Footprint}
We first examine how the CPU-resident parameter footprint affects cross-instance contention. In C2CServe, model parameters are not staged in HBM; instead, they are fetched from CPU memory over C2C on demand during inference. As a result, parameter footprint becomes a first-order determinant of C2C traffic intensity: larger models generate more off-chip parameter transfers per forward pass and therefore place greater pressure on the shared interconnect.

We use a series of Llama 3 models and evaluate different colocated model pairs to compare solo-run throughput with co-run throughput, as shown in ~\fref{fig:shapeGEMM}(a). We quantify cross-instance interference as the gap between the sum of solo-run throughput and the co-run throughput. As the total size of the two colocated models increases from 5\,GB to 44\,GB, throughput decreases in both the solo-run and co-run settings. However, the co-run case degrades much more sharply than the solo baseline, and the interference gap widens from 28\% to 42\%. This is because larger colocated models generate more concurrent parameter-fetch traffic over C2C, increasing the likelihood of overlapping fetch streams across MIG instances. The resulting contention on the shared interconnect leads to substantially greater throughput degradation under co-run execution. 

This trend shows that larger parameter footprints create greater pressure on the shared C2C link, making model footprint a useful signal for scheduling. An intelligent scheduler should therefore avoid co-locating models with large CPU-resident parameters on the same GPU, and instead use footprint-aware placement to reduce C2C contention and improve aggregate throughput.

\subsubsection{Impact of Execution Granularity}
We next study how execution granularity affects cross-instance contention. In chunk-based execution, each chunk contains a group of prompt tokens whose activations {reside in HBM} and are computed against CPU-resident weights during prefill~\cite{agrawal2024taming}. Thus, chunk size does not affect where prompts are stored; instead, it determines how much reuse each C2C-fetched weight tile receives and how long each MIG instance occupies the shared C2C link.

Figure~\textcolor{blue}{\ref{fig:shapeGEMM}}(b) illustrates this tradeoff. At a small chunk size of 512, concurrent execution nearly matches the solo baseline, with only a 2\% interference gap. As chunk size increases, solo-run throughput continues to improve, indicating higher single-instance efficiency. In contrast, co-run throughput improves much more slowly, and the interference gap widens to 42\% at the full-chunk setting.

The underlying reason is that larger chunks amortize each CPU-resident weight fetch {across more HBM-resident prompt activations}, improving per-instance efficiency. However, they also create longer C2C streaming phases, occupying the shared interconnect for longer periods and increasing overlap across colocated instances. Thus, chunk size is not only a local efficiency knob; it also controls the C2C interference imposed on neighboring instances.

\textbf{Takeaway.} 
To mitigate shared C2C contention, the scheduler jointly optimizes placement and execution granularity. It avoids colocating models with high parameter-fetch demand and dynamically adjust chunk size under contention to smooth C2C traffic across MIG instances.

\section{Overview of C2CServe Architecture}

\begin{figure}[tbp]	\centerline{\includegraphics[width=0.9\linewidth]{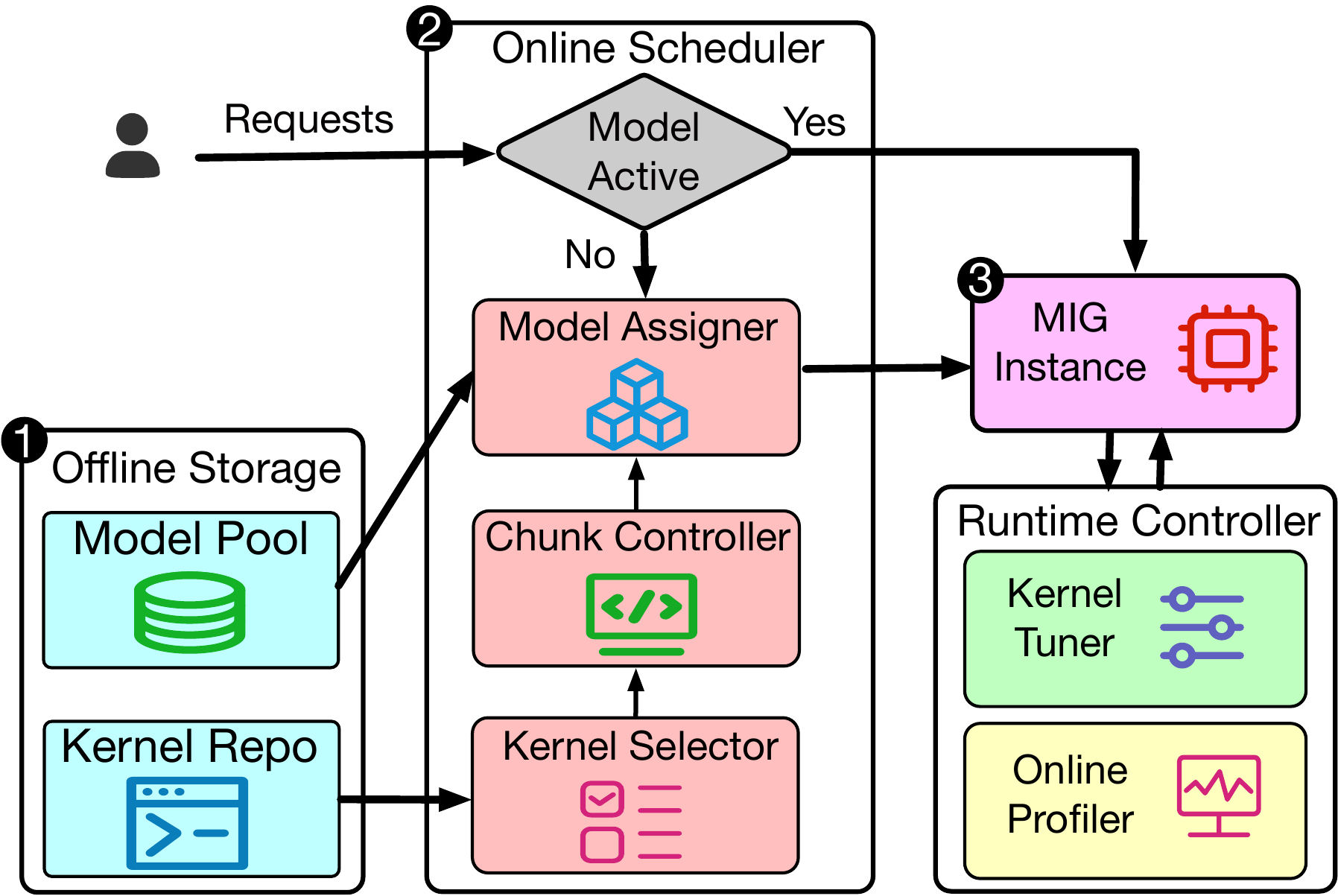}}
	\vspace{-0.5em}
    \caption{System architecture of C2CServe.}
    \vspace{-0.15in}
	\label{fig:systemArchi}
\end{figure}

C2CServe is a Superchip-native serverless LLM serving system, with the overall architecture shown in Fig.~\ref{fig:systemArchi}. For each incoming request with an SLO requirement, C2CServe schedules the request to a suitable MIG instance and selects the corresponding execution configuration including the model placement, chunk size and kernel knob
for high performance.
\textsc{C2CServe} is a Superchip-native serverless LLM serving system, as shown in Fig.~\ref{fig:systemArchi}. For each request, \textsc{C2CServe} selects the MIG instance, model placement, chunk size, and HybridGEMM kernel knob for high performance.

The \textit{Offline Storage} {\color[RGB]{0, 0, 0}{\ding{182}}} maintains a CPU-resident \textit{Model Pool} and an offline \textit{Kernel Repository}. The model pool stores multiple LLMs in host memory to reduce HBM residency pressure, while the kernel repository provides pre-compiled kernel variants for different parameter precisions, layer shapes, and MIG configurations.

The \textit{Online Scheduler} {\color[RGB]{0, 0, 0}{\ding{183}}} coordinates three decisions: the \textit{Model Assigner} places models onto MIG instances according to model footprint and expected C2C traffic; the \textit{Chunk Controller} adjusts chunk size to trade off GPU efficiency and C2C burstiness; and the \textit{Kernel Selector} chooses the appropriate kernel.

The incoming request is then executed on a MIG Instance, where activations and KV cache remain in HBM while model weights can be streamed from CPU memory over NVLink-C2C. The \textit{Runtime Controller} {\color[RGB]{0, 0, 0}{\ding{184}}} closes the feedback loop: the \textit{Online Profiler} monitors latency, HBM usage, C2C bandwidth, and cross-instance interference, while the \textit{Kernel Tuner} refines kernel choices and tuning parameters. This runtime feedback enables C2CServe to adapt scheduling and kernel execution under dynamic multi-model workloads.

\section{HybridGEMM Kernel Design}
\label{sec:HybridGEMM-Kernel}
This section introduces {HybridGEMM} as shown in Algorithm~\ref{alg:HybridGEMM}, a GEMM kernel that adapts to the dynamic bandwidth gap between CPU--GPU interconnects and HBM on MIG-enabled Superchips. 

\begin{algorithm}[t]
\DontPrintSemicolon
\footnotesize
\captionsetup[algorithm]{justification=raggedright,singlelinecheck=false}
\caption{HybridGEMM with split ratio $\alpha\!\in\![0,1]$.
$W\!\in\!\mathbb{R}^{K\times N}$, $X\!\in\!\mathbb{R}^{M\times K}$, $O\!\in\!\mathbb{R}^{M\times N}$.}
\label{alg:HybridGEMM}

\SetKwProg{Procedure}{Procedure}{:}{}
\SetKwFunction{FMixed}{\textbf{HybridGEMM}}
\SetKwFunction{FSym}{\textbf{SymmetricKernel}}
\SetKwFunction{FAsym}{\textbf{AsymmetricKernel}}

\Procedure{\FMixed{$X$, $W$, $O$, $\alpha$}}{
  $N_{\text{sym}} \gets \lfloor \alpha \cdot N \rfloor$;\quad
  \label{step:HybridGEMM}
  $N_{\text{asym}} \gets N - N_{\text{sym}}$\;
  Partition $W = [\,W_{\text{sym}}\!\mid\!W_{\text{asym}}\,]$ and
  $O = [\,O_{\text{sym}}\!\mid\!O_{\text{asym}}\,]$ along $N$\;
  \textcolor{blue}{\tcc*[l]{SM-level parallelism}}
  \FSym{$X,\,W_{\text{sym}},\,O_{\text{sym}}$}\;
  \FAsym{$X,\,W_{\text{asym}},\,O_{\text{asym}}$}\;
  \label{step:HybridGEMMEnd}
}

\vspace{0.25em}

\Procedure{\FSym{$X,\,W,\,O$}}{
  \label{step:SymKernel}
  \textcolor{blue}{\tcc*[l]{SM owns one output tile $O_{m,n}$ per iteration}}
  Allocate SM shared-memory buffers $\mathsf{sX}$, $\mathsf{sW}$\;
  Allocate register accumulator $\mathsf{acc}$\;

  \For{$k \gets 1$ \KwTo $T_K$}{
    \label{step:LoadX}
    $\mathsf{TMA.Load}(\mathsf{sX}, X_{m,k})$\;
    \label{step:LoadW}
    $\mathsf{TMA.Load}(\mathsf{sW}, W_{n,k})$\;
    \textcolor{blue}{\tcc*[l]{Compute the tile using tensor-core MMA}}
    $\mathsf{WMMA}(\mathsf{acc}, \mathsf{sX}, \mathsf{sW})$\;
  }

  \textcolor{blue}{\tcc*[l]{Write accumulated output tile back to HBM}}
  $\mathsf{TMA.Store}(O_{m,n}, \mathsf{acc})$\;
  \label{step:SymKernelEnd}
}

\vspace{0.25em}

\Procedure{\FAsym{$X,\,W,\,O$}}{
  \label{step:AsymKernel}
  Allocate SM shared-memory buffers $\mathsf{sX}$, $\mathsf{sW}$\;
  Allocate register $\mathsf{tmp}$\;

  \textcolor{blue}{\tcc*[l]{Each SM pins one weight tile $W_{n,k}$}}
  $\mathsf{sW} \gets \mathsf{TMA.Load}(W_{n,k})$\;
  \label{step:AsymKernelsWLoad}

  \For{$m \gets 1$ \KwTo $T_M$}{
    \label{step:AsymKernelIteraction}
    $\mathsf{sX} \gets \mathsf{TMA.Load}(X_{m,k})$\;
    $\mathsf{WMMA}(\mathsf{tmp}, \mathsf{sX}, \mathsf{sW})$\;
    \textcolor{blue}{\tcc*[l]{TMA fuses accumulation and HBM writeback}}
    \label{step:AsymKernelsTMAReduction}
    $\mathsf{TMA.Reduction}(O_{m,n}, \mathsf{tmp})$\;
  }
}
\end{algorithm}

\subsection{Existing Kernels Fall Short}

GEMM kernels commonly use tiling to partition large matrices into smaller blocks that fit within the GPU memory hierarchy, such as shared memory and registers. By loading each tile once and reusing it across multiple multiply--accumulate operations, tiling reduces redundant memory traffic and improves compute efficiency. 

For GEMM \(O=XW\), tiling strategies mainly differ in which tiles remain stationary during computation. 
A conventional \emph{symmetric GEMM} follows an output-stationary dataflow as shown in~\fref{fig:HybridGEMM}(a): it keeps an \(O\) tile in registers and repeatedly updates it while loading the corresponding \(X\) and \(W\) tiles into shared memory. These operand tiles are brought from HBM to shared memory (\sref{step:LoadX} and \textcolor{blue}{\ref{step:LoadW}}) using Tensor Memory Accelerator (TMA), a hardware engine for asynchronous tiled transfers, and are then consumed by tensor cores for matrix multiply--accumulate operations (\sref{step:SymKernel}). In contrast, an \emph{asymmetric GEMM} is weight-stationary as shown in~\fref{fig:HybridGEMM}(b): it keeps a CPU-resident \(W\) tile in shared memory~(\sref{step:AsymKernelsWLoad}) and reuses it across multiple \(M\)-dimension iterations~(\sref{step:AsymKernelIteraction}), reducing repeated CPU--GPU transfers. This dataflow, however, introduces additional HBM overhead because partial outputs are accumulated in HBM-resident \(O\) tiles rather than entirely in registers. The asymmetric kernel mitigates this overhead with TMA reduction~(\sref{step:AsymKernelsTMAReduction}), which fuses partial accumulation with HBM writeback. When \(X\) and \(O\) reside in HBM while \(W\) resides in CPU memory, the two designs are complementary: symmetric tiling is HBM-frugal but C2C-fragile because it repeatedly fetches CPU-resident weights, whereas asymmetric tiling is C2C-frugal but HBM-heavy because it accumulates partial outputs repeatedly in HBM.

Neither tiling strategy is uniformly optimal on MIG-enabled Superchips. MIG partitioning reshapes the HBM--C2C bandwidth balance: HBM bandwidth is partitioned across MIG instances, whereas C2C bandwidth remains shared. With more MIG instances, per-instance HBM bandwidth drops and the HBM--C2C gap narrows, favoring symmetric GEMM. With fewer MIG instances, per-instance HBM bandwidth is more abundant, making C2C the bottleneck; in this regime, asymmetric GEMM is preferable because it reduces repeated CPU--GPU weight transfers.

\subsection{Efficient HybridGEMM}
\subsubsection{Tiling Pattern Analysis}
The above tradeoff makes a fixed tiling choice fragile: a pattern tuned for one MIG partition or contention level can become suboptimal under another. HybridGEMM avoids this fixed choice by exposing a continuous knob that controls the fraction of work assigned to the symmetric and asymmetric paths, balancing HBM and C2C traffic under the current MIG configuration~(\sref{step:HybridGEMM}).

This knob is necessary because the two paths stress different resources. The symmetric path preserves the optimized conventional GEMM structure but is sensitive to repeated C2C weight fetches. The asymmetric path reduces C2C traffic by increasing weight reuse, but shifts overhead to HBM-side partial-output accumulation. HybridGEMM therefore selects an execution mix that matches the current HBM--C2C bandwidth balance.

\subsubsection{Runtime-Tunable HybridGEMM}
\label{sec:Hybrid_Execution}
Rather than committing to a single tiling pattern, HybridGEMM exposes a split between SymGEMM and AsymGEMM, as shown in~\fref{fig:HybridGEMM}(c). Specifically, HybridGEMM partitions the output along the $N$ dimension using a ratio $\alpha\in[0,1]$ (\sref{step:HybridGEMM}): the first region of width $N_{\mathrm{sym}}=\lfloor\alpha N\rfloor$ uses the symmetric path, while the remaining region of width $N_{\mathrm{asym}}=N-N_{\mathrm{sym}}$ uses the asymmetric path. The two paths run on different SMs and write disjoint columns of $O$, so they require no inter-stream synchronization. This allows the HBM-sensitive and C2C-sensitive execution paths to run concurrently and overlap their respective stalls. Importantly, neither path stages $W$ into HBM; weights remain CPU-resident and are streamed on demand, making the optimization transparent to the user and requiring no explicit weight management.

The optimal $\alpha$ is runtime-dependent. Beyond workload shape and MIG partitioning, it must account for live C2C contention from co-resident tenants. As multiple MIG instances stream CPU-resident weights concurrently, each instance's effective C2C bandwidth changes over time, making a static $\alpha$ fragile. {C2CServe} therefore treats $\alpha$ as a runtime tuning knob; its selection is handled by the online tuning module (detailed in~\cref{sec:Online_Fine_tuning}).

\begin{figure}[tbp]	\centerline{\includegraphics[width=0.95\linewidth]{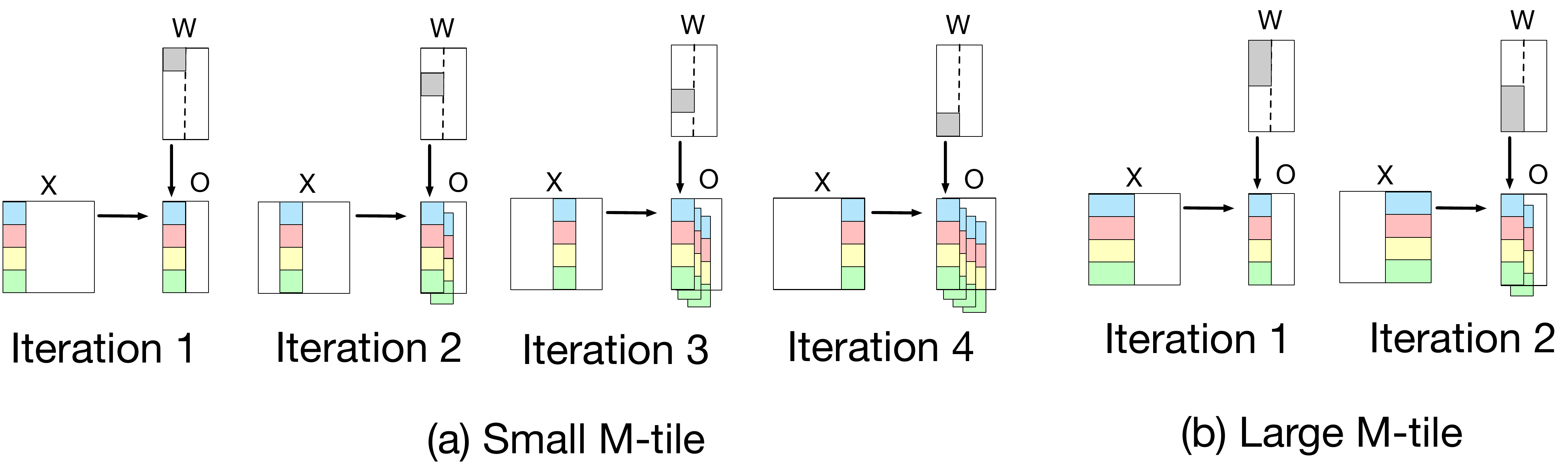}}
	\vspace{-0.5em}
    \caption{Impact of \(M\)-dimension tile size in asymmetric GEMM.}
    \vspace{-0.15in}
	\label{fig:asym_tile_m}
\end{figure}

\subsubsection{Tiling Tradeoff in HybridGEMM}
The two execution paths in HybridGEMM prefer different tiling choices. The symmetric path follows conventional GEMM tiling and performs well when the cache hierarchy provides sufficient reuse of streamed weight tiles~\cite{CUTLASS}. 

The asymmetric path requires a different tiling strategy. It reduces C2C traffic by reusing a CPU-resident \(W\) tile in shared memory, but shifts overhead to HBM because partial outputs must be repeatedly accumulated in HBM. To reduce this HBM overhead, the asymmetric path favors a larger output tile along the \(M\) dimension, as illustrated in ~\fref{fig:asym_tile_m}. A larger \(M\)-tile allows each loaded \(W\) tile to be reused across more rows of \(X\) before partial outputs are written back, reducing the number of HBM accumulation steps. This trades additional on-chip resource usage for fewer HBM transactions. HybridGEMM therefore exposes flexible tiling choices to balance shared-memory usage, HBM traffic, and C2C traffic under the current MIG configuration.
\section{Contention-aware Online Scheduling}
\label{sec:scheduling}
In this section, we introduce the policy used by the Online Scheduler module ({\color[RGB]{0, 0, 0}{\ding{183}}} in Fig.~\ref{fig:systemArchi}) for serving CPU-resident models on MIG-partitioned Superchips. The scheduler jointly manages shared C2C bandwidth, per-MIG HBM bandwidth, and TTFT/TPOT 
targets through three decisions: model placement, chunk-size selection, and HybridGEMM kernel selection.

\subsection{Scheduling Workflow}
\label{sec:scheduling_workflow}
For each incoming request, the scheduler performs four steps, as shown in Fig.~\ref{fig:systemArchi}. First, it checks whether the requested model is already active on a MIG instance. If so, the request is directly routed to that instance, avoiding inference-engine initialization and model setup overheads. Second, if the model is not active, the scheduler decides where to place it. If an idle MIG instance is available, the scheduler chooses the placement that fits the MIG-local HBM bandwidth budget {while respecting the aggregate C2C bandwidth budget shared by all MIG instances on the Superchip}. Otherwise, it triggers model switching by evicting an inactive or low-priority model instance and reusing the released MIG partition for the new model. Third, after placement, the scheduler selects the request chunk size using the offline profiling table. The selected chunk size must satisfy the TTFT/TPOT target while keeping the estimated HBM demand below the MIG-local bandwidth limit.
This decision is made per model and per MIG size (detailed in~\tref{tab:gh200_mig_config}),
since different MIG partitions provide different compute capacity and HBM bandwidth. Finally, the scheduler selects an initial HybridGEMM ratio from the profiling table and refines it online using feedback from measured latency, C2C utilization, and HBM utilization. 

\subsection{Bandwidth-Aware Model Placement}
At initialization, the scheduler first determines which CPU-resident models can be served concurrently on MIG instances without oversubscribing the shared NVLink-C2C link. Since model weights remain in pinned CPU memory, C2CServe treats each active model as a C2C bandwidth consumer rather than an HBM-capacity consumer.

For each model \(m\), the scheduler estimates the C2C bandwidth required to meet its decode target \(TPOT_m\). Since model weights are streamed from CPU memory during execution, weight-fetch latency forms a lower bound on the achievable per-token latency. In contrast, TTFT depends on the execution time of the scheduled prefill chunk, which we handle through MIG-aware chunk sizing (\cref{sec:MIG_Aware_Chunk}). Given the CPU-resident weight footprint \(S_m\) and target \(TPOT_m\), the required C2C bandwidth is estimated as:

\[
BW^{C2C}_m = \frac{S_m}{TPOT_m},
\] where the weight footprint \(S_m\) can be obtained from the model configuration.

The scheduler uses this estimate as a feasibility constraint during both initial active-set construction and request-driven model switching described in \cref{sec:scheduling_workflow}. Specifically, the active model set \(M\) is valid only if its aggregate C2C demand fits within the available link bandwidth:
\[
\sum_{m \in M} BW^{C2C}_m \leq BW^{C2C}_{avail}.
\]
This constraint prevents C2C oversubscription across MIG instances. Under-packing leaves the shared C2C link idle and reduces model concurrency, while over-packing causes C2C contention and degrades TTFT/TPOT.

\subsection{MIG-Aware Chunk Sizing}
\label{sec:MIG_Aware_Chunk}
After assigning a model to a MIG instance, the scheduler selects the prefill chunk size for that instance, which mainly affect TTFT. Larger chunks improve tensor-core utilization and reduce scheduling overhead, but increase activation traffic and output accumulation pressure within the MIG partition. Smaller chunks reduce per-step HBM pressure, but may underutilize compute resources and require more execution steps to complete prefill.

Since the parameter-heavy computation in LLM inference is dominated by GEMMs, especially in MLP layers, we use a representative operation \(O=XW\) to illustrate chunk-size selection, where \(X\) and \(O\) reside in GPU memory and \(W\) resides in pinned CPU memory. For each candidate chunk size, the scheduler estimates the HBM bandwidth demand required to satisfy the TTFT target
\(L_{\text{TTFT}}\):
\[
BW^{HBM} =
\frac{\gamma_X \cdot S_X + \gamma_O \cdot S_O}{L_{\text{TTFT}}},
\]
where \(S_X\) is the activation size and \(S_O\) is the output size for the candidate chunk. The coefficients \(\gamma_X\) and \(\gamma_O\) capture the effective number of HBM reads/writes induced by the selected HybridGEMM dataflow. They depend on the GEMM tiling shape and are calibrated through offline profiling. Intuitively, \(\gamma_X \cdot S_X\) captures the activation traffic inside the MIG instance, while \(\gamma_O \cdot S_O\) captures the output traffic and accumulation traffic introduced by the selected dataflow. 

Chunk-size selection is constrained by both MIG-local HBM bandwidth and compute capacity. Offline, C2CServe profiles candidate chunk sizes for each model and MIG configuration, and records the smallest chunk size that satisfies the TTFT target without exceeding the instance's HBM and compute budgets. At runtime, the scheduler looks up this profiling table to select the chunk size for each request. C2CServe then uses online feedback control to adjust the HybridGEMM ratio under bandwidth contention, keeping TTFT and TPOT within their required ranges.

\subsection{Kernel Selection}
During initialization, \textsc{C2CServe} selects a set of candidate kernels based on the model configuration, including model precision, tensor shape, and the target MIG partition. Since different precisions and tile sizes require different kernel variants, the runtime first maps each model to a prebuilt HybridGEMM kernel family that matches its execution format.

For each selected kernel family, \textsc{C2CServe} initializes the HybridGEMM ratio \(\alpha\) to 0, favoring the C2C-frugal dataflow to minimize NVLink-C2C contention when a model is first activated. This conservative starting point is important because the C2C interconnect is shared across MIG instances, and aggressive CPU-memory streaming can interfere with co-located models. Once requests arrive, \textsc{C2CServe} tunes \(\alpha\) online using observed HBM bandwidth, C2C bandwidth, and latency: it shifts toward a more C2C-efficient dataflow when C2C is saturated, and toward a more HBM-efficient dataflow when HBM becomes the bottleneck. Thus, kernel selection provides a safe initial configuration, while runtime feedback adapts HybridGEMM to the available heterogeneous-memory bandwidth.
\section{Online Fine-tuning}
\label{sec:Online_Fine_tuning}
To control under C2C contention, it will be better to fine-tune the kernel to meet the performance requirement. 

\subsection{Why Static Tuning Fails}
As discussed in~\cref{sec:Hybrid_Execution}, HybridGEMM uses a tuning knob \(\alpha\) to balance work between \emph{SymGEMM} and \emph{AsymGEMM}. A static \(\alpha\) assumes stable bandwidth for each MIG slice, but NVLink-C2C is shared across all instances. Thus, each tenant's effective C2C bandwidth changes with co-resident models. An offline-optimal \(\alpha\) can therefore become suboptimal at runtime. C2CServe tunes \(\alpha\) online, providing a lightweight control mechanism to adapt performance without changing the request chunk size.

\subsection{Runtime Signals for HybridGEMM Control}
\label{sec:alpha_signals}
HybridGEMM exposes a control knob \(\alpha\) (\cref{sec:Hybrid_Execution}) that sets the fraction of output columns executed by the symmetric GEMM path. Larger \(\alpha\) reduces HBM-side output accumulation but increases repeated accesses to CPU-resident weights, while smaller \(\alpha\) improves CPU-weight reuse but increases HBM traffic for partial-output updates. Since the HBM--C2C bandwidth balance varies across MIG profiles and co-tenant activity, C2CServe tunes \(\alpha\) online with lightweight feedback control. 

The controller is applied to parameter-heavy GEMM operators in each transformer layer, such as the MLP projections, where CPU-resident weights dominate C2C traffic. For each GEMM, C2CServe assigns a latency budget \(L_{\mathrm{budget}}\) by distributing the request-level TTFT or TPOT target across the profiled execution time of the model's GEMM operators. At runtime, the controller periodically observes the measured GEMM latency \(L\), HBM bandwidth utilization \(U_{\mathrm{HBM}}\), and NVLink-C2C bandwidth utilization \(U_{\mathrm{C2C}}\). The bandwidth utilizations are normalized by the available bandwidth of the corresponding MIG profile and Superchip link:
\[
U_{\mathrm{HBM}} =
\frac{BW_{\mathrm{measured}}^{\mathrm{HBM}}}
     {BW_{\mathrm{avail}}^{\mathrm{HBM}}},
\qquad
U_{\mathrm{C2C}} =
\frac{BW_{\mathrm{measured}}^{\mathrm{C2C}}}
     {BW_{\mathrm{avail}}^{\mathrm{C2C}}}.
\]

\subsection{Feedback Update for HybridGEMM Tuning}
\label{sec:alpha_update}
Algorithm~\ref{alg:alpha_feedback} summarizes the feedback control loop. At each control interval, C2CServe computes the bandwidth imbalance:
\[
\Delta = U_{\mathrm{C2C}} - U_{\mathrm{HBM}} .
\]
If \(\Delta > \tau\), C2C is more saturated and the controller decreases \(\alpha\), shifting more work to \emph{AsymGEMM}. If \(\Delta < -\tau\), HBM is more saturated and the controller increases \(\alpha\), shifting more work to \emph{SymGEMM}. Otherwise, \(\alpha\) remains unchanged. The update is:
\[
\alpha_{t+1} =
\mathrm{clip}_{[0,1]}
\left(
\alpha_t - \eta_t \cdot \mathrm{sign}(\Delta)
\right),
\]
where \(\tau\) is the imbalance threshold, \(\eta_t\) is a latency-aware step size, and \(\mathrm{clip}_{[0,1]}(\cdot)\) bounds \(\alpha\) within \([0,1]\).

To avoid oscillation, C2CServe smooths latency and bandwidth measurements with an exponential moving average, updates \(\alpha\) only when \(|\Delta|>\tau\), and bounds the maximum step size per interval. The controller runs at layer or request boundaries for each active MIG instance, avoiding changes inside an in-flight kernel. Since \(\alpha\) only changes the selected prebuilt HybridGEMM variant or the SM partition ratio, the controller does not modify chunk size, model placement, or weight layout, keeping runtime tuning lightweight.

Specifically, at each control interval, C2CServefirst smooths the measured latency and bandwidth utilization to filter out short-term noise.
It then compares normalized C2C and HBM utilization to identify the dominant bottleneck. If the two paths are balanced, the controller keeps \(\alpha\) unchanged. Otherwise, it adjusts \(\alpha\) toward the less-contended dataflow, using a larger step when the GEMM latency exceeds its budget and a smaller step when the operator already meets the target.

\begin{algorithm}[t]
\footnotesize
\caption{C2CServe feedback control for HybridGEMM ratio}
\label{alg:alpha_feedback}
\KwIn{Current ratio \(\alpha_t\), latency \(L\), target \(L_{\mathrm{slo}}\), HBM utilization \(U_{\mathrm{HBM}}\), C2C utilization \(U_{\mathrm{C2C}}\)}
\KwOut{Updated ratio \(\alpha_{t+1}\)}

Smooth \(L\), \(U_{\mathrm{HBM}}\), and \(U_{\mathrm{C2C}}\) using EMA\;
\(\Delta \gets U_{\mathrm{C2C}} - U_{\mathrm{HBM}}\)\;

\If{\(|\Delta| < \tau\)}{
  \(\alpha_{t+1} \gets \alpha_t\)\;
}
\Else{
  \eIf{\(L > L_{\mathrm{slo}}\)}{
    \(\eta \gets \eta_{\mathrm{fast}}\)\;
  }{
    \(\eta \gets \eta_{\mathrm{slow}}\)\;
  }

  \eIf{\(\Delta > 0\)}{
    \textcolor{blue}{\tcp{C2C is more saturated; shift toward AsymGEMM}}
    \(\alpha_{t+1} \gets \max(0, \alpha_t - \eta)\)\;
  }{
    \textcolor{blue}{\tcp{HBM is more saturated; shift toward SymGEMM}}
    \(\alpha_{t+1} \gets \min(1, \alpha_t + \eta)\)\;
  }
}
\Return \(\alpha_{t+1}\)\;
\end{algorithm}
\section{Implementation}
\label{sec:impl}
C2CServe is implemented with approximately 5K lines of CUDA/C++ and 4K lines of Python, with a fork delta of about 3K lines against upstream Mini-SGLang~\cite{mini-sglang}. CPU-resident weights are allocated using \texttt{cudaHostAlloc}~\cite{cudahostalloc} with the \texttt{cudaHostAllocMapped} flag, which provides pinned and device-mapped host memory. Kernels access these weights through device pointers. These accesses use ordinary global-memory load instructions and are transparently routed over C2C through the unified address space; no specialized intrinsics or driver-level mechanisms are required.

We integrate C2CServe into a fork of Mini-SGLang~\cite{mini-sglang}. The HybridGEMM dispatcher replaces Mini-SGLang's cuBLAS calls in the projection operators of attention and MLP layers, while the C2CServe scheduler replaces Mini-SGLang's request batcher. KV-cache management, continuous batching, and the API frontend remain unchanged. 





\section{Evaluation}
\label{sec:eval}

\subsection{Experimental Setup}
\label{sec:eval_setup}
\textbf{Hardware.}
We evaluate C2CServe on an NVIDIA GH200 Grace Hopper Superchip with 480\,GB Grace LPDDR5X memory, 96\,GB HBM3, and C2C interconnect with up to \(\sim\)900\,GB/s peak bandwidth. We evaluate both full-GPU execution and partitioned MIG execution. The MIG configurations are summarized in~\tref{tab:gh200_mig_config}.

\textbf{Software.} Our evaluation server runs Ubuntu 22.04.5 LTS on an aarch64 Linux platform. We use CUDA {12.8}~\cite{cuda}, PyTorch {2.7.0}~\cite{pytorch} and CUTLASS {4.3.5}~\cite{CUTLASS}. C2CServe is implemented on top of mini-sglang~\cite{mini-sglang}.

\textbf{Workloads.}
We use GenTD26~\cite{lin2025understanding,GenAI}, an anonymized production generative-AI serving trace from Alibaba that contains 3.5 million requests across 87 models over three weeks on 300+ GPUs. We use ShareGPT~\cite{sharegpt} to derive prompt and output length distributions for latency microbenchmarks.

\textbf{Baselines.}
\begin{itemize}\itemsep0pt
\item \textbf{ServerlessLLM (SLLM)}~\cite{fu2024serverlessllm}: Reduces LLM cold-start latency with multi-tier checkpoint loading, live migration, and locality-aware scheduling.

\item \textbf{Aegaeon}~\cite{xiang2025aegaeon}: Enables multi-model GPU pooling through token-granularity scheduling and auto-scaling with low scaling overhead.

\item \textbf{MoE-Infinity}~\cite{moe-infinity}: Exploits sparse expert reuse in batch-one MoE inference to guide expert caching and prefetching on memory-limited machines.

\item \textbf{FineMoE}~\cite{yu2026taming}: Reduces MoE memory usage via fine-grained expert offloading guided by expert-selection patterns and prompt hints.
\end{itemize}

\textbf{Models.} We evaluate dense and MoE models~\cite{huggingfaceDataset} according to the scope of each baseline. For dense-model baselines, including ServerlessLLM and Aegaeon, we use Llama-3~\cite{dubey2024llama} models with Llama-3B, Llama-8B and Llama-70B parameters. For MoE baselines, including ServerlessLLM, MoE-Infinity, and FineMoE, we use Mixtral-8x7B~\cite{jiang2024mixtral} and Qwen3-30B-A3B~\cite{yang2025qwen3}. All models use BF16 precision.

\textbf{Performance Metrics.} 
We measure LLM serving performance using the 95th-percentile TTFT and TPOT, and evaluate serving elasticity using cold-start latency and model-switch latency.

\begin{figure}[t]
  \centering
  \includegraphics[width=\columnwidth]{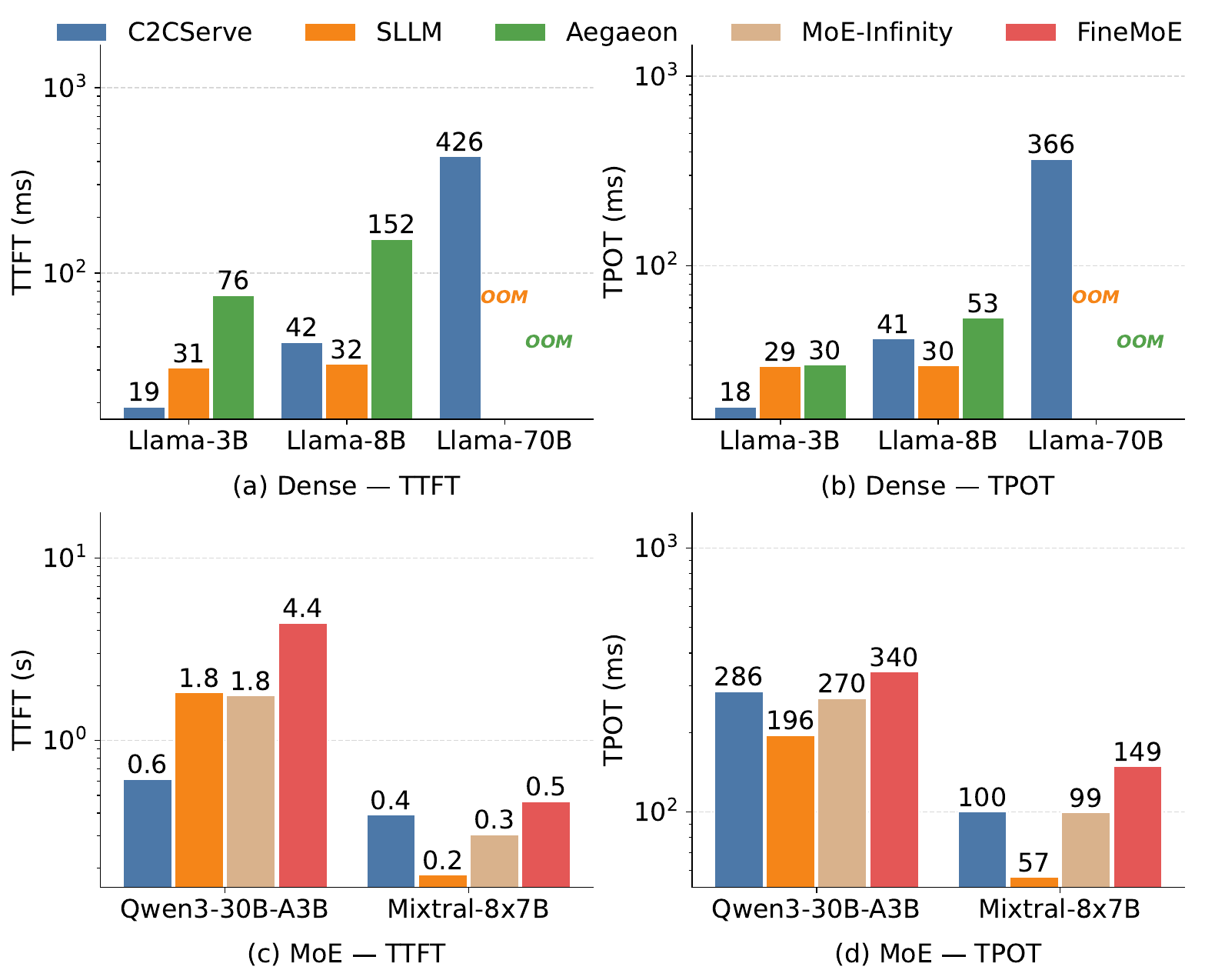}
  	\vspace{-1.5em}
  \caption{TTFT and TPOT comparison across baselines.}
  	\vspace{-0.5em}
  \label{fig:steady_throughput}
\end{figure}

\label{sec:eval_trace}

    \subsection{End-to-End Evaluation}
    \subsubsection{Full-GPU Serving Performance}
    We evaluate serving performance on the full GPU, as shown in~\fref{fig:steady_throughput}. For dense models, C2CServe is competitive with ServerlessLLM on the commonly supported models, achieving the same average TPOT and slightly lower average TTFT. Compared with Aegaeon, C2CServe reduces TTFT and TPOT by 3.8\(\times\) and 1.5\(\times\) on average, respectively. In addition, C2CServe enables Llama-70B execution, where both ServerlessLLM and Aegaeon run out of memory.

    For MoE models, C2CServe provides substantial TTFT reduction on Qwen3-30B-A3B, improving over ServerlessLLM, MoE-Infinity, and FineMoE by 3.0\(\times\), 2.9\(\times\), and 7.2\(\times\), respectively. On Mixtral-8x7B, ServerlessLLM achieves the lowest TTFT, while C2CServe still outperforms FineMoE. For TPOT, C2CServe remains competitive with MoE-specific baselines: it nearly matches MoE-Infinity on Mixtral-8x7B and reduces TPOT over FineMoE by 1.2--1.5\(\times\), though ServerlessLLM achieves lower TPOT on both MoE models.

    Overall, these results show that shifting model residency from HBM to CPU memory primarily removes cold-start and capacity bottlenecks, while HybridGEMM keeps the runtime streaming overhead low enough to preserve competitive per-token performance.
    
    \subsubsection{Cold-Start Overhead under MIG Partitioning}
    We measure end-to-end cold-start latency when the requested model is not resident in any active inference engine. This includes loading model parameters from disk and initializing the inference runtime from scratch, with latency measured from request arrival to the generation of the first token.

\begin{figure}[t]
      \centering
      \includegraphics[width=\columnwidth]{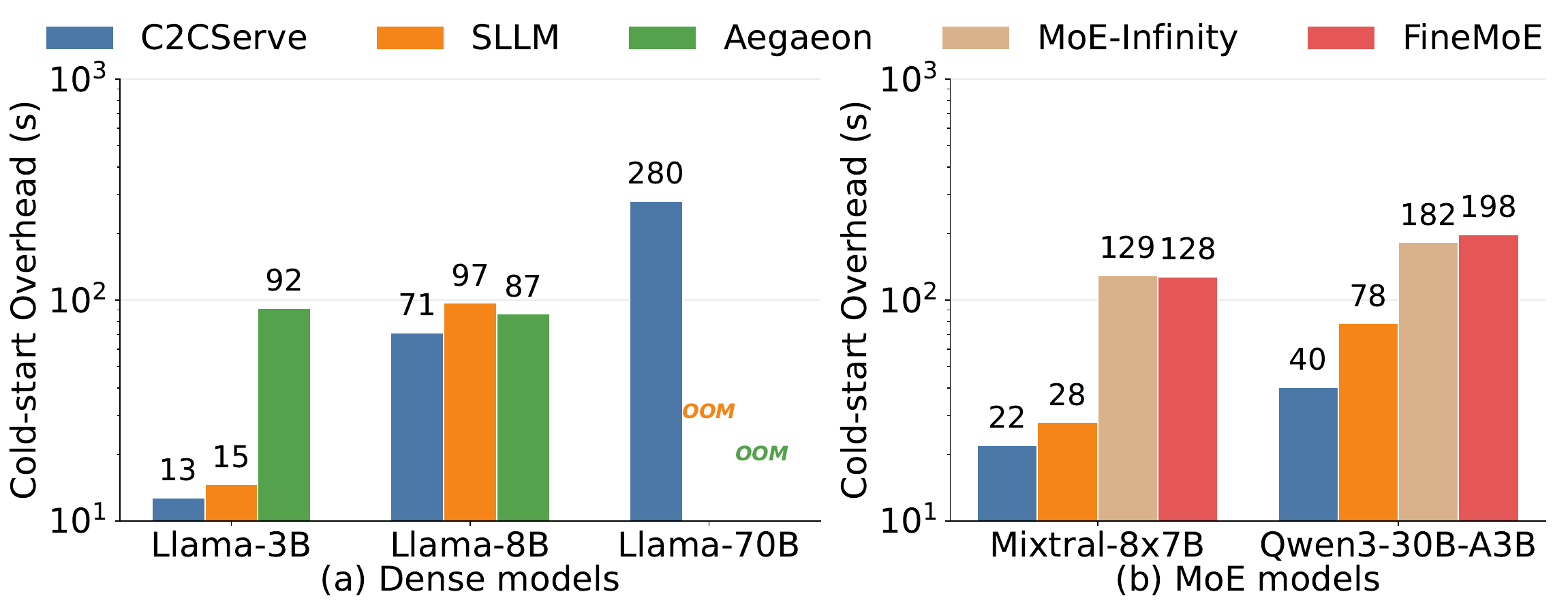}
      	\vspace{-2.em}
      \caption{Cold-start latency}
      \label{fig:cold_start}
    \end{figure}

Figure~\textcolor{blue}{\ref{fig:cold_start}} compares cold-start latency across dense and MoE models. For dense models, C2CServe reduces latency over ServerlessLLM by 1.15--1.37\(\times\) on commonly supported models and enables Llama-70B, where both ServerlessLLM and Aegaeon run out of memory. Compared with Aegaeon, C2CServe improves latency by up to 7.1\(\times\). For MoE models, C2CServe reduces cold-start latency over MoE-Infinity and FineMoE by 4.6--5.0\(\times\), and outperforms ServerlessLLM by 1.95\(\times\) on Qwen3-30B-A3B. Overall, C2CServe avoids HBM-capacity failures while maintaining low cold-start latency across dense and MoE workloads.
    
    \subsubsection{Model Switch under MIG Partitioning}
\begin{figure}[t]
      \centering
      \includegraphics[width=\columnwidth]{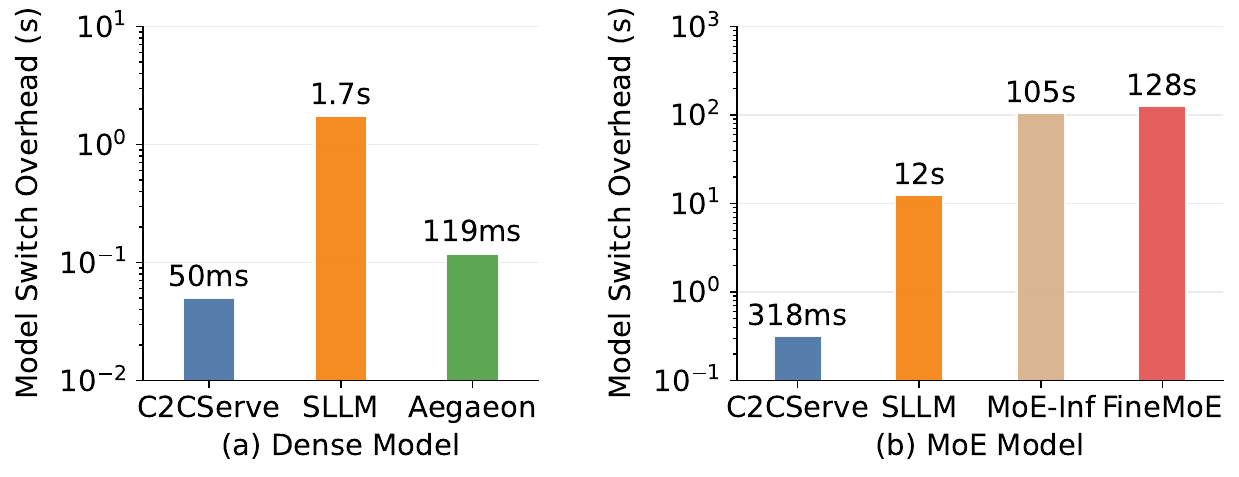}
      	\vspace{-1.5em}
          \caption{Model-switch Overhead.}
        	\vspace{-0.5em}
      \label{fig:model_switching}
    \end{figure}

    We evaluate warm-runtime model-switch latency in~\fref{fig:model_switching}, where the model has been loading into pinned CPU memory rather the in the disk before the switching.

    Compared with existing systems, C2CServe reduces model-switch overhead by one to three orders of magnitude. On dense models, it lowers switching overhead from 1.7\,s in ServerlessLLM and 119\,ms in Aegaeon to 50\,ms. On MoE models, it reduces switching overhead from 12\,s in ServerlessLLM, 105\,s in MoE-Infinity, and 128\,s in FineMoE to 318\,ms. These results show that CPU-resident model weights turn model switching from a heavyweight reload path into a lightweight runtime operation.        
    
\begin{figure}[t]
  \centering
  \includegraphics[width=\columnwidth]{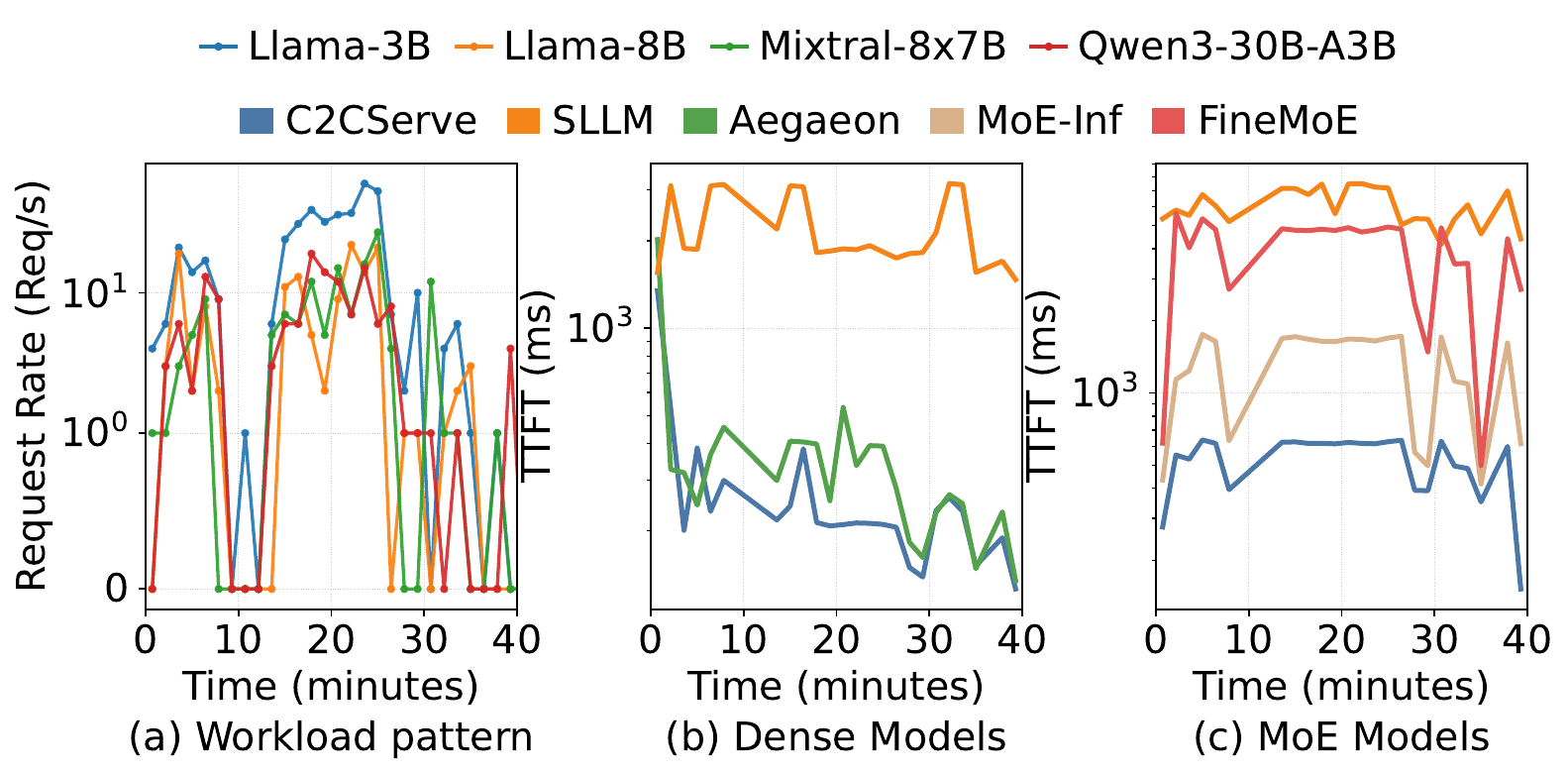}
  	\vspace{-1.5em}
  \caption{MoE and Dense Models trace replay.}
    	\vspace{-0.5em}
  \label{fig:dynamic_workload}
\end{figure}

\subsection{Dynamic Workload}
We replay a production-derived dynamic workload using open-source models, as shown in~\fref{fig:dynamic_workload}(a). The trace is bursty and time-varying, stressing model switching and latency stability. Figures~\textcolor{blue}{\ref{fig:dynamic_workload}}(b) and~\textcolor{blue}{~\ref{fig:dynamic_workload}}(c) report TTFT over time. On dense models,C2CServe keeps TTFT around 0.2--0.7\,s, compared with 2--7\,s for ServerlessLLM and 0.3--2\,s for Aegaeon. On MoE models,C2CServe stays around 0.5--0.8\,s, while ServerlessLLM, MoE-Infinity, and FineMoE often reach several seconds and occasionally exceed 10\,s.
Under a 1\,s TTFT target, C2CServe meets the latency requirement for 95\% of requests. These results show that CPU-resident weights reduce model-switching overhead and queueing pressure under bursty demand.

\subsection{Component-Level Evaluation}
\label{sec:eval_components}
In this section, we will study how each component in C2CServe contribute to the end-to-end performance.

\subsubsection{Benefits of HybridGEMM}
\begin{figure}[t]
  \centering
  \includegraphics[width=0.7\columnwidth]{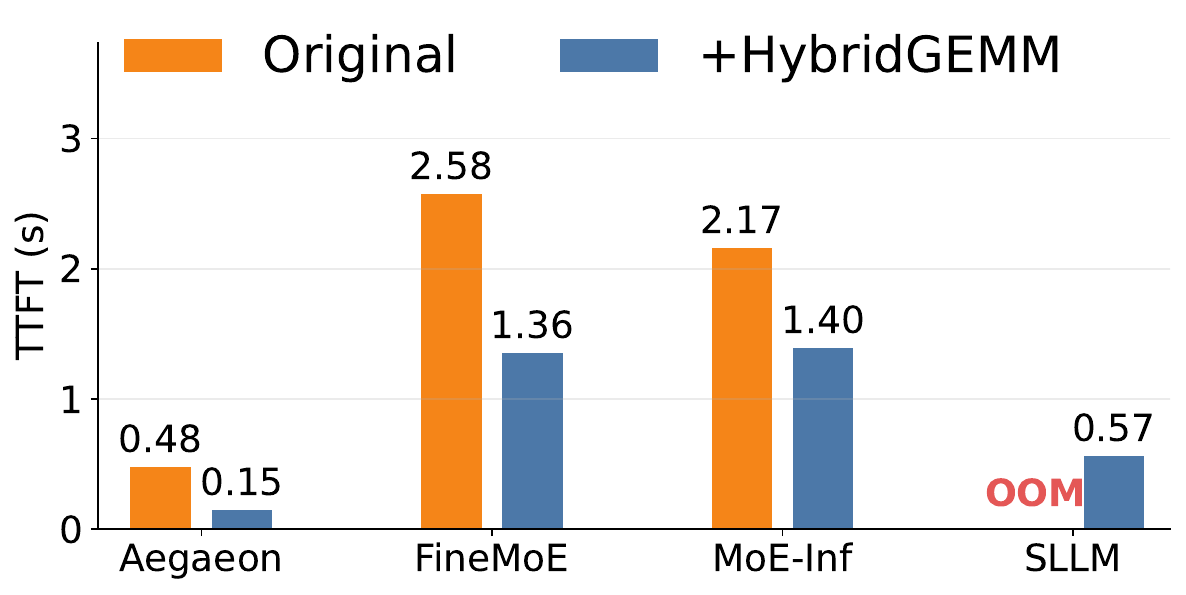}
  	\vspace{-0.5em}
  \caption{Baseline integrated with HybridGEMM.}
    	\vspace{-0.5em}
  \label{fig:retrofit}
\end{figure}

The key advantage of C2CServe is that it keeps model parameters in CPU memory and accesses them directly during execution, avoiding expensive GPU weight loading on the cold-start path. To examine whether existing systems can benefit from this mechanism alone, we integrate HybridGEMM into several baselines, as shown in~\fref{fig:retrofit}. With HybridGEMM, Aegaeon reduces TTFT from 0.48\,s to 0.15\,s, a 68\% reduction. HybridGEMM also reduces TTFT from 2.58\,s to 1.36\,s for FineMoE and from 2.17\,s to 1.40\,s for MoE-Infinity, corresponding to 47\% and 36\% reductions, respectively. In addition, HybridGEMM enables ServerlessLLM to serve Llama-70B, which otherwise runs out of HBM, achieving 0.57\,s TTFT. These results show that direct CPU-memory execution can substantially reduce weight-loading overhead while enabling larger models to be served.

\begin{figure}[t]
  \centering
  \includegraphics[width=0.9\columnwidth]{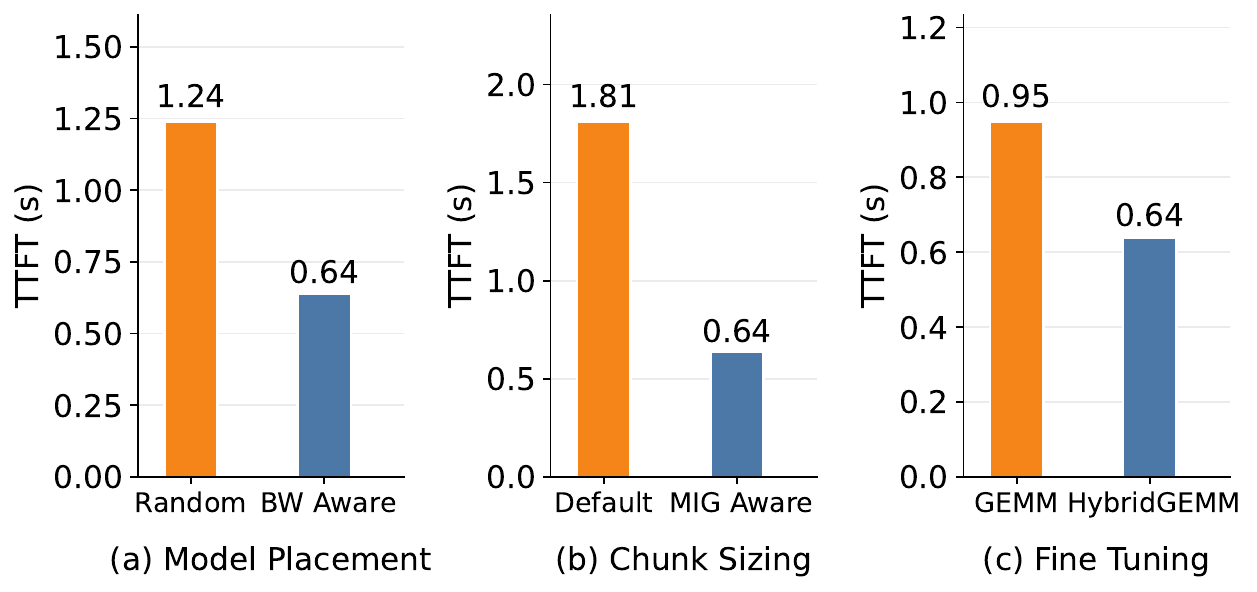}
  	\vspace{-0.5em}
  \caption{Component-level comparison.}
    	\vspace{-1.em}
  \label{fig:ablation}
\end{figure}

\subsubsection{Bandwidth-aware Placement}
We evaluate the effectiveness of bandwidth-aware placement by comparing C2CServe's smart scheduler with a random placement policy. As shown in~\fref{fig:ablation}(a), random placement results in a p99 TTFT of 1.24\,s, because mismatched model--MIG assignments can overload either the shared C2C link or the MIG-local HBM bandwidth. In contrast, the smart scheduler matches each model to a MIG slice whose remaining HBM and C2C bandwidth budgets fit its runtime demand, reducing p99 TTFT to 0.64\,s, a 1.94\(\times\) improvement. This benefit appears even when the chunk controller and HybridGEMM are already active, showing that bandwidth-aware placement is essential for controlling tail latency under multi-tenant MIG execution.

\subsubsection{Chunk-size Control}
We evaluate the effectiveness of the chunk controller by comparing the default chunk configuration with C2CServe's tuned chunk size. As shown in~\fref{fig:ablation}(b), the default configuration leads to a p99 TTFT of 1.81\,s, while the tuned chunk controller reduces it to 0.64\,s, a 2.83\(\times\) improvement. The key mechanism is that chunking controls the per-step C2C burst size: without proper tuning, peak bandwidth demand spikes during prefill, causing SLO-bound requests on small MIG slices to miss their latency targets. By selecting chunk sizes that fit the available C2C and MIG-local HBM bandwidth budgets, C2CServe smooths bandwidth demand and substantially reduces tail latency.

\subsubsection{HybridGEMM Knob Tuning}
We evaluate the effectiveness of tuning the HybridGEMM dataflow knob \(\alpha\) by comparing a static GEMM configuration with C2CServe's HybridGEMM configuration. As shown in~\fref{fig:ablation}(c), the static configuration results in a p99 TTFT of 0.95\,s, while HybridGEMM reduces it to 0.64\,s, a 1.48\(\times\) improvement. This improvement comes from adapting the GEMM dataflow to the runtime bandwidth balance: a fixed configuration can overload either C2C or HBM bandwidth, whereas HybridGEMM balances C2C and HBM pressure across MIG slices. This result shows that HybridGEMM must be tuned jointly with placement and chunk sizing rather than configured statically.

\subsection{Projection to Future Superchips}
We use a simple hardware-projection model to examine whether C2CServe follows the Superchip hardware trend. As summarized in Table~\ref{tab:superchip_hw}, we use GH200~\cite{GH200} as the baseline and assume the GEMM data-access pattern remains unchanged across generations. The key tradeoff is that conventional HBM-resident GEMM mainly benefits from higher HBM bandwidth, whereas HybridGEMM is primarily bounded by C2C bandwidth when streaming CPU-resident weights. On GB200~\cite{GB200}, CPU memory and C2C bandwidth remain similar to GH200, while HBM bandwidth increases by about \(2\times\), so conventional GEMM benefits more from the hardware upgrade than HybridGEMM.

For Rubin-class hardware~\cite{rubin}, the trend becomes more favorable to C2CServe: CPU memory, HBM bandwidth, and C2C bandwidth increase by about \(3.2\times\), \(5.5\times\), and \(2\times\) over GH200, respectively. Although HybridGEMM scales with C2C bandwidth and therefore improves more slowly than HBM-resident GEMM, the \(3.2\times\) larger CPU memory tier can expand the CPU-resident model pool enough to offset the relative GEMM performance gap. This suggests that C2CServe is aligned with the Superchip roadmap, where CPU memory becomes a larger active weight tier that GPUs can directly access through high-bandwidth C2C links.

\begin{table}[t]

\centering

\scriptsize

\setlength{\tabcolsep}{3pt}
\caption{Superchip hardware information.}
\label{tab:superchip_hw}
\begin{tabular}{c|c|c|c|c}
\toprule
Platform & CPU mem. & HBM & HBM BW & C2C BW \\
\midrule
GH200~\cite{GH200} & 480\,GB & 96\,GB & 4.0\,TB/s & 900\,GB/s \\
GB200~\cite{GB200} & 480\,GB & 192\,GB & 8.0\,TB/s & 900\,GB/s \\
Rubin~\cite{rubin} & 1.5\,TB & 288\,GB & 22\,TB/s & 1.8\,TB/s \\
\bottomrule
\end{tabular}
\end{table}
\section{Related Work}
\label{sec:related_work}
\textbf{Serverless LLM serving.}
Recent systems improve LLM serving elasticity by reducing cold-start overhead,
improving model multiplexing, or offloading model states. ServerlessLLM~\cite{fu2024serverlessllm}
uses multi-tier checkpoint loading, live migration, and locality-aware
scheduling to reduce cold starts, while Aegaeon~\cite{xiang2025aegaeon}
improves multi-model GPU pooling with token-granularity scheduling and
autoscaling. Medusa~\cite{zeng2025medusa} and Foundry~\cite{liu2026foundry}
reduce startup overhead by materializing CUDA graph states offline and restoring
them online. For MoE serving, MoE-Infinity~\cite{moe-infinity} and
FineMoE~\cite{yu2026taming} reduce GPU memory pressure through expert caching,
prefetching, and offloading. These systems optimize model loading, runtime
state reconstruction, or expert placement, but generally rely on staging weights
or experts into GPU memory before computation. C2CServe instead keeps
weights in CPU memory and executes over them directly through NVLink-C2C,
enabling request-granularity model switching on MIG instances without reloading
weights into HBM.

\textbf{GPU resource multiplexing.}
GPU sharing improves utilization by multiplexing workloads in time or space.
Temporal multiplexing~\cite{vGPUs} shares GPU time through context switching,
whereas spatial sharing allows concurrent execution on the same GPU to reduce
switching overhead~\cite{vijaykumar2016zorua}. NVIDIA MIG~\cite{MIGs} provides
hardware-supported spatial sharing by partitioning a GPU into isolated instances
with dedicated compute and memory resources. Orion~\cite{strati2024orion}
improves fine-grained spatial sharing by co-scheduling operators according to
compute and memory demand, REEF~\cite{han2022microsecond} improves temporal
sharing with microsecond-scale kernel preemption and controlled concurrency, and
LLMStation~\cite{he2025resource} combines spatial and temporal multiplexing for
concurrent LLM fine-tuning and inference. Unlike these systems, C2CServe
targets MIG-based LLM serving on CPU--GPU Superchips, where the key challenge is
coordinating MIG-local HBM resources with the shared NVLink-C2C bandwidth used
to stream CPU-resident weights.

\textbf{Superchip-based LLM systems.}
CPU--GPU Superchips such as NVIDIA GH200 have motivated new designs for LLM
offloading and memory management. SuperOffload~\cite{lian2026superoffload}
optimizes LLM training by jointly using the GPU, Grace CPU, and NVLink-C2C.
For inference, SuperInfer~\cite{yu2026superinfer} improves TTFT SLO attainment
with SLO-aware request rotation and full-duplex NVLink-C2C KV transfer,
Pie~\cite{xu2024pie} expands effective memory with performance-transparent
CPU--GPU swapping, and Oneiros~\cite{li2025oneiros} remaps memory from inactive
model parameters to KV cache to reduce swapping. C2CServe is
complementary: rather than optimizing training offload, KV transfer, or memory
swapping, it targets MIG-based serverless serving and uses NVLink-C2C to execute
directly over CPU-resident model weights across isolated MIG instances.

\section{Conclusion}
\label{sec:conclusion}
We present C2CServe, a request-granularity serverless LLM serving system that enables MIG instances to switch models across requests without loading weights into HBM. C2CServe combines HybridGEMM, a heterogeneous-memory GEMM kernel that balances HBM and NVLink-C2C traffic, with a hierarchical scheduler that jointly controls model placement, chunk sizing, and runtime kernel selection.

\bibliographystyle{plain}
\bibliography{serverlessLLM}

\end{document}